\pdfoutput=1

\documentclass[11pt,twoside,a4paper,cmspaper,final,collab]{cms-tdr}
\def\svnVersion{473639P}\def\cmsCernNoTag{CERN-EP-2018-187}\def\cmsCernDate{\today}\def\cmsMessage{Published in the European Physical Journal C as \href{http://dx.doi.org/10.1140/epjc/s10052-018-6144-y}{\doi{10.1140/epjc/s10052-018-6144-y.}}}

\begin{document}\cmsNoteHeader{FSQ-16-011}

\hyphenation{had-ron-i-za-tion}
\hyphenation{cal-or-i-me-ter}
\hyphenation{de-vices}
\RCS$HeadURL: svn+ssh://jgradosl@svn.cern.ch/reps/tdr2/notes/FSQ-16-011/trunk/FSQ-16-011.tex $
\RCS$Id: FSQ-16-011.tex 422420 2017-08-25 12:12:34Z jgradosl $

\newcommand{\QGS}{\textsc{qgsjet}II\xspace}
\newcommand{\LHC}{LHC\xspace}
\newcommand{\EPOS}{\textsc{epos}\xspace}
\newcommand{\CUETPM}{\textsc{cuetp8m1}\xspace}
\newcommand{\CUETPS}{\textsc{cuetp8s1}\xspace}
\newcommand{\CUETM}{\textsc{cuetm1}\xspace}
\newcommand{\MBR}{\textsc{mbr4c}\xspace}
\newcommand{\fourC}{\textsc{4c}\xspace}
\newcommand{\pse}{pseudorapidity }

\newlength\cmsTabSkip\setlength{\cmsTabSkip}{1ex}

\newenvironment{tolerant}[1]{\par\tolerance=#1\relax}{ \par }

\newcommand{\ptz}{\ensuremath{p_{\perp 0}}\xspace}
\newcommand{\ptmin}{\ensuremath{p_{\text{T,min}}}\xspace}
\newcommand{\ptlead}{\ensuremath{p_{\text{T,leading}}}\xspace}
\newcommand{\Eta}{\ensuremath{\eta}\xspace}
\newcommand{\nch}{\ensuremath{N_{\text{ch}}}\xspace}

\newlength\cmsFigWidth
\ifthenelse{\boolean{cms@external}}{\setlength\cmsFigWidth{1\columnwidth}}{\setlength\cmsFigWidth{0.473\textwidth}}
\ifthenelse{\boolean{cms@external}}{\providecommand{\cmsLeft}{top\xspace}}{\providecommand{\cmsLeft}{left\xspace}}
\ifthenelse{\boolean{cms@external}}{\providecommand{\cmsRight}{bottom\xspace}}{\providecommand{\cmsRight}{right\xspace}}
\cmsNoteHeader{FSQ-16-011}
\providecommand{\cmsTable}[1]{\resizebox{\textwidth}{!}{#1}}

\title{Measurement of charged particle spectra in minimum-bias events from proton-proton collisions at $\sqrt{s}=13\TeV$}

\date{\today}

\abstract{Pseudorapidity, transverse momentum, and multiplicity distributions
 are measured in the pseudorapidity range $\abs{\eta} < 2.4$ for charged particles with transverse momenta  satisfying $\pt > 0.5\GeV$ in proton-proton collisions at a center-of-mass energy of $\sqrt{s} = 13\TeV$.
Measurements are presented in three different event categories. The most inclusive of the categories corresponds to an inelastic \Pp \Pp\ data set, while the other two categories are exclusive subsets of the inelastic sample that are either enhanced or depleted in single diffractive dissociation events.
The measurements are compared to predictions from Monte Carlo event generators used to describe
high-energy hadronic interactions in collider and cosmic-ray physics.
}

\hypersetup{%
pdfauthor={CMS Collaboration},%
pdftitle={Measurement of charged particles spectra in proton-proton collisions at sqrt{s}=13~TeV with the CMS experiment},%
pdfsubject={CMS},%
pdfkeywords={CMS, physics, software, computing}}

\maketitle

\hyphenation{multi-plicity multi-plicities NSD-en-hanced}

\section{Introduction}
The study of the properties of particle production without any selection bias arising from requiring
the presence of a hard scattering (a selection known as "minimum bias")
is one of the most basic measurements that can be made at hadron colliders.
Such events are produced by strong interactions of partons inside the hadrons, which occur at low momentum exchanges, for which  predictions of quantum chromodynamics (QCD) cannot be obtained perturbatively,
and for which diffractive scatterings or multiple partonic interactions (MPI) play a significant role.
The theoretical description of these components of particle production is based on phenomenological
models with free parameters adjusted ("tuned") to reproduce the experimental data.
However, when a momentum transfer of several GeV (referred to as a hard process) is involved, predictions obtained from perturbative QCD (pQCD) are, in many cases, in good agreement with the measurements.
Understanding the transition region between hard processes calculable with perturbative techniques  and soft processes described by nonperturbative models is required for a full description of particle production in proton-proton (\Pp \Pp ) collisions at the LHC.
It is also essential when the collider is operated at high luminosities since a bunch crossing
contains many \Pp \Pp\ collisions (pileup) forming a complex final state that needs to be theoretically controlled for precise studies of
standard model processes, as well as for new physics searches.

Inclusive measurements of charged particle \pse distributions, $\rd{}\nch/\rd{}\eta$, and transverse momentum distributions, $\rd{}\nch/\rd{}\pt$, as well as charged particle multiplicities have been previously performed in proton-proton  and proton-antiproton collisions
in the center-of-mass energy range $\sqrt{s} =  $0.2--8\TeV
and in various phase space regions~\cite{UA5, UA4_dndeta, UA1, CDF:1990, CDF:2009, ALICE2010a, Aamodt:2010pp,Adam:2015gka,ALICE:2017pcy,Aad:2010rd,ATLAS2011, Aad:2016xww, CMSdndeta1, CMSdndeta2, Khachatryan:2010nk,CMS-PAPERS-FSQ-12-026,Khachatryan:2015fia,LHCb2011, TOTEM2012_dndeta}.
Most of these measurements are described to within 10--20\% by present event generators
as reported, \eg, in Ref.~\cite{Khachatryan:2015pea}.

More recently, measurements of the charged-hadron pseudorapidity distribution in \Pp \Pp\  collisions at the highest energies reached so far, $\sqrt{s} = 13 \TeV$, have been presented in Refs.~\cite{Adam:2015pza,Aad:2016mok, Aaboud:2016itf,Khachatryan:2015jna}. The present work extends those studies, for charged particles with $\PT > 0.5\GeV$ measured over the range   \hbox{$\abs{\eta} < 2.4$}, to cover not only the pseudorapidity density, but also the per-event multiplicity probability, $P(\nch)$, as well as different transverse-momentum distributions, such as that of the leading charged particle $\rd{}\nch/\rd{}\pt$, and its corresponding integrated spectrum, $D(\ptmin)$.
The integrated spectrum $D(\ptmin)$ is defined as:
\begin{equation}
D(\ptmin)
= \frac{1}{N_{\text{events}}}  \int_{\ptmin} \rd\ptlead \left( \dd{N}{\ptlead}\right) .
\label{int}
\end{equation}
Here $N_{\text{events}}$ is the number of selected events, $N$ is the number of events with a leading charged particle with transverse momentum \ptlead, and \ptmin is the lower limit of the integral.
In each event, the highest-\PT charged particle within $\abs{\eta} < 2.4 $ and with $\PT  > 0.5 \GeV$ is selected as the leading charged particle.
The integrated spectrum of charged particles is sensitive to the transition between the nonperturbative and perturbative QCD regions~\cite{Grebenyuk:2012qp,Khachatryan:2015fia}.

 \begin{tolerant}{800}
The measured distributions are presented for three different event data sets:  an inelastic sample, a sample dominated by nonsingle diffractive dissociation events (NSD-enhanced sample), and a sample enriched by single diffractive dissociation events (SD-enhanced sample).
The measurements are compared to predictions from different Monte Carlo (MC) event generators
used to describe high-energy hadronic interactions in collider and cosmic-ray physics.
\end{tolerant}

This article is organized as follows.
Section~\ref{sec:CMS} gives a brief description of the CMS detector.
The MC models used for corrections and comparison to data are described in Section~\ref{sec:mc}.
The data sample, track reconstruction, and event selection are discussed in Section~\ref{sec:data}.
The procedure to correct the data for detector effects and the systematic uncertainties affecting the measurements are described in Sections~\ref{sec:corrections} and \ref{sec:sys}, respectively.
The final results are presented in Section~\ref{sec:res} and a summary is given in Section~\ref{sec:sum}.

\section{The CMS detector}
\label{sec:CMS}
The central feature of the CMS apparatus is a superconducting solenoid of 6\unit{m} internal diameter, providing a magnetic field of 3.8\unit{T}. Within the solenoid volume are a silicon pixel and strip tracker, a lead tungstate crystal electromagnetic calorimeter, and a brass and scintillator hadron calorimeter, each composed of a barrel and two endcap sections. Forward calorimeters extend the pseudorapidity coverage provided by the barrel and endcap detectors. Muons are detected in gas-ionization chambers embedded in the steel flux-return yoke outside the solenoid.

The tracking detector consists of 1440 silicon pixel and 15\,148 silicon strip detector modules.
The barrel is composed of 3 pixel and 10 strip layers around the primary interaction point (IP) at radii ranging from 4.4 to 110\unit{cm}.
The forward and backward endcaps each consist of 2 pixel disks and 12 strip disks in up to 9 rings.
Three of the strip rings and four of the barrel strip layers contain an additional plane, with a stereo angle of 100\unit{mrad}, to provide a measurement of the $r$- and $z$-coordinate, respectively.
The silicon tracker measures charged particles within the \pse range $\abs{\eta}< 2.5$.
For particles of $1 < \pt < 10\GeV$ and $\abs{\eta} < 1.4$, the track resolutions are typically 1.5\% in \pt and 25--90 (45--150)\mum in the transverse (longitudinal) impact parameter~\cite{CMS-PAPERS-TRK-11-001}.

The hadron forward (HF) calorimeter uses steel as an absorber and quartz fibers as the sensitive material. The HF calorimeters are located  at 11.2\unit{m} from the interaction region, one on each end, and together they provide coverage in the range $2.9 < \abs{\eta} < 5.2$. Each HF calorimeter consist of 432 readout towers, containing long and short quartz fibers running parallel to the beam. The long fibers run the entire depth of the HF calorimeter (165\unit{cm}, or approximately 10 interaction lengths), while the short fibers start at a depth of 22\unit{cm} from the front of the detector. By reading out the two sets of fibers separately, it is possible to distinguish showers generated by electrons and photons, which deposit a large fraction of their energy in the long-fiber calorimeter segment, from those generated by hadrons, which produce on average nearly equal signals in both calorimeter segments. Calorimeter towers are formed by grouping bundles of fibers of the same type.
Bundles of long fibers form the electromagnetic towers and bundles of short fibers form the hadronic towers.

A more detailed description of the CMS detector, together with a definition of the coordinate system used and the relevant kinematic variables, can be found in Ref.~\cite{Chatrchyan:2008zzk}.

\section{Theoretical predictions}
\label{sec:mc}
Three  different event generators simulating  hadronic collisions are used to correct the measurements to particle level (Section~\ref{sec:corrections}), and for comparisons with the final results.
Simulated event samples were used to optimize the event selection, vertex selection, and tracking efficiencies.

The \PYTHIA{}8 (version 8.153) event generator \cite{sjostrand2008}  uses a model \cite{sjostrand2008,skands2007}
in which initial-state radiation and multiple partonic interactions are interleaved.
Parton showers in \PYTHIA are modeled according to the Dokshitzer--Gribov--Lipatov--Altarelli--Parisi (DGLAP) evolution equations~\cite{Gribov:1972ri, Altarelli:1977zs, Dokshitzer:1977sg}, and hadronization  is based on the Lund string fragmentation model \cite{andersson:1983}.
Diffractive cross sections are described by the Schuler--Sj{\"o}strand model \cite{Sch:94}.
Particle production from a low-mass state X, with $M_\mathrm{X} < 10\GeV$, is described by the Lund string fragmentation model, while for higher masses, $M_\mathrm{X} > 10\GeV$, a perturbative description of the pomeron-proton scattering is introduced.
The latter is based on diffractive parton distribution functions~\cite{H12006, H12007, diffractivePDFs_2013}, which represent probability distributions for partons inside the proton, under the constraint that the proton emerges intact from the collision.
The \PYTHIA{}8 generator is used with the tune \CUETPM{}~\cite{Khachatryan:2015pea}
(also referred to as \CUETM{}), which is based on the Monash tune~\cite{Skands:2014pea} using the NNPDF2.3LO~\cite{Ball:2011uy,Ball:2013hta}
parton distribution function (PDF) set, with parameters optimized to reproduce underlying event (UE) data from CMS at $\sqrt{s} = 7 \TeV$ and CDF at $\sqrt{s} = 1.96 \TeV$.

The minimum bias Rockefeller (MBR) model~\cite{Ciesielski:2012mc}  is also implemented within the \PYTHIA{}8 event generator.  When used in conjunction with the \fourC tune~\cite{Corke:2010yf}  (which includes parton showering) it is referred to as the \PYTHIA{}8 \MBR  model.
This model reproduces the measured energy dependence
of the total, elastic, and inelastic \Pp\Pp\ cross sections, and can be used to fully simulate the main diffractive components of the inelastic cross section.
The generation of diffractive processes is based on a phenomenological renormalized Regge model~\cite{Goulianos:2002vm, Goulianos:2004as},  interpreting the pomeron flux as the probability of forming a diffractive rapidity gap.
The value of the pomeron intercept $\alpha(0)=1.08$ is found to give the best description of the diffractive dissociation cross sections measured by CMS at $\sqrt{s}=7\TeV$~\cite{Khachatryan:2015gka}.

The data are also compared to predictions from the \EPOS\ ~\cite{werner2006} MC event generator (version 1.99) used in cosmic ray physics \cite{denterria:2011}, including contributions from soft- and hard-parton dynamics.
The soft component is described in terms of the exchange of virtual quasi-particle states, as in Gribov's Reggeon field theory \cite{gribov1968}, with multi-pomeron exchanges accounting for UE effects.
At higher energies, the interaction is described in terms of the same
degrees of freedom (reggeons and pomerons), but generalized to include hard processes via hard-pomeron scattering diagrams, which are equivalent to a leading order pQCD approach with DGLAP evolution.
The \EPOS\ generator is used with the \LHC tune \cite{Pierog:2013ria, Pierog:2013}.

 \begin{tolerant}{800}
Event samples obtained from the event generators \PYTHIA{}8, \PYTHIA{}8 \textsc{mbr}, and \EPOS are passed through the CMS detector simulation based on \GEANTfour~\cite{GEANT4}, and are processed and reconstructed in the same manner as collision data.
The number of pileup interactions in the MC samples is adjusted
to match the distribution in the data.
\end{tolerant}

\section{Data set, track reconstruction, and event selection}
\label{sec:data}
In order to minimize the effect of pileup, the data considered in the analysis were collected in a special run in summer 2015 with an average number of \Pp\Pp\  interactions per bunch crossing of 1.3~\cite{Khachatryan:2015lva}.

The two LHC beam position monitors closest to the IP for each LHC experiment,
called the beam pick-up timing experiment (BPTX) detectors, are used to trigger the detector readout.
They are located around the beam pipe at a distance of 175\unit{m} from the IP on either side, and are designed to provide precise information on the bunch structure and timing of the incoming beams.
Events are selected by requiring the presence of both beams crossing at the IP, as inferred from the BPTX detectors.

The CMS track reconstruction algorithm is based on a combinatorial track finder (CTF)~\cite{tracking}.
The collection of reconstructed tracks is obtained through multiple iterations of the CTF reconstruction sequence.
The iterative tracking sequence consists of six iterations.
The first iteration is designed to reconstruct  prompt tracks (originating near the \Pp\Pp\ interaction point) with three pixel hits and $\pt > 0.8\GeV$.
The subsequent iterations are intended to recover prompt tracks that only have two pixel hits or lower \pt.
At each iteration an extrapolation of the trajectory  is performed, and using the Kalman filter, additional strip hits compatible with the trajectory are assigned.

High-purity tracks~\cite{CMS-PAPERS-TRK-11-001} are selected with a reconstructed   $\pt > 0.5\GeV$, in order to have high tracking efficiency (${>}80\%$) and a relative transverse momentum uncertainty smaller than 10\%.
Tracks are measured within the \pse range $ \abs{\eta} < 2.4$ corresponding to the fiducial acceptance of the tracker, in order to avoid effects from tracks very close
to its geometric edge at $\abs{\eta}=2.5$.
The impact parameter with respect to the beam spot in the transverse plane, $d_{xy}$, is required to satisfy $\abs{d_{xy}/\sigma_{xy}} < 3$, while for the point of closest approach to the primary vertex along the $z$-direction, ($d_{z}$), the requirement $\abs{d_z/\sigma_z} < 3$ is imposed. Here $\sigma_{xy} $ and $\sigma_z$ denote the uncertainties in $d_{xy}$ and $d_{z}$, respectively.
The number of pixel detector hits associated with a track has to be at least 3 in the  \pse region $\abs{\eta} \leq 1$ and at least 2 for $\abs{\eta} > 1$.

Rejection of beam background events and events with more than one collision per bunch crossing is achieved by requiring exactly one reconstructed primary vertex~\cite{CMS-PAPERS-TRK-11-001}.
The vertex produced by each collision is required to be within $\abs{z} < 15\cm$ with respect to the center of the luminous region along the beamline and within $0.2\cm$ in the transverse direction.

Different event classes are defined based on activity in the HF calorimeters by requiring the presence of at least one tower with an energy above the threshold value of 5\GeV in the fiducial acceptance region, $3 < \abs{\eta} < 5$.
The veto condition is defined by an energy deposit in the towers less than a given threshold value. An inelastic sample consists of events with activity on at least  one side of the calorimeters, whereas an NSD-enhanced sample contains those with calorimeter activity on both sides.
An SD-enhanced sample is defined by requiring activity on only one side of the calorimeters, with a veto condition being applied to the other side.

 \begin{tolerant}{800}
The HF energy threshold of 5\GeV was determined from the measurement of electronic noise and beam-induced background in the HF calorimeters, using an event sample for which a single beam was circulating in the LHC ring, and an event sample without beams.
The threshold of 5\GeV keeps the background due to noise low, while still maintaining a high selection efficiency.
The fraction of events with at least one HF tower on either side of the detector with an energy above the threshold of 5\GeV in the event samples with no collisions (one beam or no beam) is 0.13\%.
The efficiency of the event selection defined by the presence of at least one tower with an energy above 5\GeV in either side of the calorimeters is 99.3\%, and is calculated with respect to the event sample defined by the presence of exactly one reconstructed primary vertex.
\end{tolerant}

In total  a sample of 2.23 million events is selected containing 2 million NSD events and 0.23 million SD-enhanced events.

\section{Correction to particle level}
\label{sec:corrections}
The data are corrected for tracking and event selection efficiencies, as well as for detector resolution effects.
The corrected distributions correspond to stable primary charged particles, which are either directly produced in \Pp\Pp\ collisions or result from decays of particles with decay length $< 1 \cm$.
At particle level, events are selected if at least one charged particle is found within $\abs{\eta} < 2.4$  and $\pt > 0.5\GeV$.
The different event selections are defined in a similar manner to how they are defined at detector level in order to avoid any bias towards a specific MC model.
Activity in the forward region is defined by the presence of at least one particle, either charged or neutral, with an energy above 5\GeV in  the region $3 < \abs{\eta} < 5$ (referred to as the trigger particle).
The veto condition is equivalently defined  by the absence of particles with an energy above 5\GeV. The inelastic data set is defined by requiring a trigger particle in the \pse range $3 < \eta < 5$ or $-5 < \eta < -3$.
The NSD-enhanced event sample is defined by requiring a trigger particle in the regions $3 < \eta < 5$ and $-5 < \eta < -3$.
The SD-enhanced event sample is defined by requiring a trigger particle in either the positive or negative \Eta range,  with the veto condition applied to the other region.
The SD-enhanced event sample is further divided into two exclusive subsets according to the \Eta region in which the trigger particle is detected.
These subsets are referred to as SD-One-Side enhanced event samples.
Table~\ref{table:stableLevel} shows a summary of the event selection definitions at particle level.
{   \newcommand{\ms}{{ side}$^-$}
   \newcommand{\ps}{{ side}$^+$}
   \begin{table*}[h]
        \topcaption{Summary of stable-particle level definitions for each of the event samples, corresponding to the inelastic, the NSD-enhanced, and SD-enhanced categories.
Charged particles are selected with $\pt > 0.5\GeV$ and  $\abs{\eta} < 2.4$.
Forward trigger particles correspond to those with energy $ E> 5\GeV$ located in  \ms\ (defined as $-5 < \eta < -3$) and/or \ps\ (defined as $3 < \eta < 5$).  Similarly, a
{veto} corresponds to the absence of a trigger particle with $ E> 5\GeV$ in  \ms\  and/or  \ps .     }
       \centering
       \begin{tabular}{cc}
       \hline
       Event sample &  Forward region energy selection \\
       \hline
       Inelastic &  Trigger particle in \ms\ or \ps \\
       NSD-enhanced       &  Trigger particle in \ms\ and \ps \\
       SD-enhanced        &  Trigger particle in \ms (\ps) and veto  in \ps (\ms )   \\
      \hline
       \end{tabular}
       \label{table:stableLevel}
   \end{table*}
}

A response matrix, $R$, is constructed using the information provided by the MC event generators and by the full detector simulation.
The elements of the response matrix ($R_{ji}$) represent the conditional probability that (for a given observable) a true value~$i$ is measured as a value~$j$.
In this analysis, two different correction procedures were implemented.
The first method (method 1) makes use of the full detector simulation, while in the second method (method 2) a parametrization of the detector response is implemented in order to overcome the statistical limitations of the full detector simulation.
Whenever it is possible to accurately parametrize the detector resolution, method 2 is used, otherwise method 1 is applied.
The two implemented methods account for unreconstructed particles (misses) and wrongly reconstructed tracks (misreconstruction), as well as for the event selection efficiency including the vertex and enhanced-event selection.
In method 1, the correction is performed in two steps.
The first step corrects for detector resolution, missed particles and misreconstructed tracks, using an unfolding procedure.
The second step corrects for the event selection efficiency.
In method 2, the $R$ matrix is constructed in a manner that does not include information on the missed particles and misreconstructed tracks;  therefore a correction factor to account for these effects is applied.
This correction factor takes into account the missed particles and the misreconstructed tracks, as well as the event selection efficiency.
In contrast to  method 1, this correction factor is applied as a function of the observable of interest.
The reason for this is that the number of missing particles in the reconstruction and the number of misreconstructed tracks both show a dependence on the different observables.
The D'Agostini method~\cite{DAGOSTINI1995487} is used to unfold the detector effects.

The final corrected distributions are obtained by taking the average of two corrected distributions, each corrected using one of the two MC models that describe best the data at detector level.
Most of the distributions are best described by \PYTHIA{}8 \CUETM and \EPOS\ \LHC.
The only exception is the \pse distribution of the SD-enhanced event sample, for which \PYTHIA{}8 \CUETM and \PYTHIA{}8 \MBR have been used.

\section{Systematic uncertainties}
\label{sec:sys}
The following sources of systematic uncertainty are taken into account: tracking efficiency, description of the pileup modeling, sensitivity to the specific value used for the energy threshold applied to the HF towers, and the dependence on the model used for the corrections.

The systematic uncertainties show almost no dependence on the \pse of the particles in the fiducial region considered here.
For the charged particle multiplicity and \PT distributions, the systematic uncertainties are dependent on the value of the measured observable.
The systematic uncertainties associated with the \PT distributions of all charged particles, the leading charged particle, and the integrated spectrum of the latter show a similar behavior.

\begin{itemize}
\item Tracking efficiency.
The systematic uncertainty due to the difference between the track reconstruction efficiency in data and simulation is $\approx$4\%. This has been obtained in Ref.~\cite{CMS-PAPERS-FSQ-12-026} at $\sqrt{s}=8\TeV$ and validated for $\sqrt{s}=13\TeV$ data
 by comparing the tracking
 performance of the data set used in this analysis  to the performance of the same data set but reconstructed with the (different) tracking conditions used in Ref.~\cite{CMS-PAPERS-FSQ-12-026}.
The tracking efficiency is estimated with a data-driven method known as ``tag-and-probe'' \cite{CMS:2011aa} by exploiting resonances decaying into two particles.

\item Pileup modeling.
The systematic uncertainty associated with the modeling of the pileup contribution
is calculated by varying the nominal selection of events with exactly one vertex to that obtained with  at least one vertex  where only the tracks associated with the vertex with the largest
sum of the squared transverse momenta
of its tracks are used.
The difference between these two selections is taken as the associated uncertainty.
For the \pse distributions, the uncertainty is estimated to be about 1\% for the inelastic event sample,  while it is about 1.5 and 0.3\% for the NSD- and SD-enhanced event samples, respectively.
An uncertainty of about 2\% on the \PT distributions for the most inclusive selection procedure is obtained, while it is smaller than 1\%  for the  SD-enhanced event sample.
The inelastic and NSD-enhanced event samples have similar uncertainties for the multiplicity distributions, first increasing at low multiplicities from 2  to 8\% and then decreasing to 0.5\% for large multiplicities.

\item Event selection with HF.
The systematic uncertainty associated with the event selection is determined by varying the threshold applied to the energy of the HF calorimeter towers, while keeping the definition of the stable-particle level unchanged.
The default value of the energy threshold applied to the HF calorimeter towers is varied from 5\GeV by $\pm1 \GeV$.

In the inelastic and NSD-enhanced event samples an uncertainty of less than 2\% for all the relevant distributions is obtained, while for the SD-enhanced sample the uncertainty increases to $\approx$6\% for the \pse distributions and varies between 1 and 15\% for the \PT distributions.

\item Model dependence.
The systematic uncertainty due to the model dependence is calculated as one half of the difference between the corrected distributions using the two MC models mentioned in Section~\ref{sec:corrections}.
For the \pse distributions, it varies between 0.1 and 1\% for the inelastic and NSD-enhanced event samples, and is about 7\% for the SD-enhanced sample.

For the transverse-momentum distributions, the most inclusive event samples have a maximum uncertainty of about 4\% at high \PT, while the SD-enhanced event sample exhibits a maximum uncertainty of 10\% around 2\GeV, decreasing at both the low and high ends of the spectrum.

For the event multiplicity distributions, the inelastic and NSD-enhanced event samples have similar uncertainties  with values up to 8\%, reaching a maximum uncertainty for low and high multiplicities, and a minimum for multiplicities between 5 and 40.

\end{itemize}
The total systematic uncertainty is obtained by adding the different sources discussed above in quadrature.
Table~\ref{table:systematics} summarizes all the contributions per observable and per event selection.
The total uncertainty is reported for each case.

Due to the statistical limitations of the multiplicity measurement in the SD-enhanced event sample, this distribution is not included in the results.

\begin{table*}[thbp]
   \centering
     \topcaption{Summary of systematic uncertainties per observable for each of the event samples.
   The observables are (presented as rows, from top to bottom) pseudorapidity, multiplicity, transverse momentum, leading transverse momentum, and the integral of the latter.
The columns, from left to right, represent the following event samples:  Inelastic, NSD-enhanced and SD-enhanced.
   For each observable the respective sources of uncertainty are listed.  These are, from top to bottom:  the tracking efficiency, the pileup modelling, the event selection and the model dependence.  The final value in each case represents the total systematic uncertainty.
   }
   \begin{tabular}{ccccc}
  \hline
   \multicolumn{5}{c}{Systematic uncertainties [\%]} \\
   Observable & Source  & Inelastic             &  NSD-enhanced     &  SD-enhanced     \\
   \hline
   \multirow{5}{*}{$\frac{\rd\nch}{\rd\eta}$}
     				  & Tracking efficiency       &  4  &  4  &  4                           \\
				  & Pileup modeling           &  1            & 1.6     & 0.3      \\
   				  & Event selection            &   $<$0.2  & 1        & 7         \\
   				  & Model dependence      &   0.8        & 0.5     & 7         \\
  				  & {Total}                       &  {4}     & {4}  & {9 } \\[\cmsTabSkip]
   \multirow{5}{*}{$P(\nch)$}
     				  & Tracking efficiency       &   4  &  4  &  \NA                 \\
  			 	  & Pileup modeling           &   0.5--8  &  0.5--8  & \NA  \\
   				  & Event selection            &   0--2  &  0--2  &  \NA   \\
   				  & Model dependence      &   0--8  &  0--8  &  \NA  \\
  				  & {Total}                       &   {4--8}  &  {4--8}  & \NA \\[\cmsTabSkip]
   \multirow{5}{*}{$\frac{\rd\nch}{\rd \pt}$}
    				  & Tracking efficiency       &   4  &  4  & 4  					\\
    			 	  & Pileup modeling           &   0--2     &  0.5--2  &  0--2    \\
  				  & Event selection            &   $<$0.2  &  1           &  1--12  \\
   				  & Model dependence      &   0--4     &  0--4     &  4--10  \\
  				  & {Total}                       &  {4}  &  {4}  & {7 -- 14}  \\[\cmsTabSkip]
   \multirow{5}{*}{$\frac{\rd\nch}{\rd{\ptlead}}$}
    				  & Tracking efficiency       &  4  &  4  & 4  					\\
   			 	  & Pileup modeling           &  0--4     &  0.5--4  &  0--4    \\
   				  & Event selection            &  $<$0.2  &  1           &  1--12  \\
   				  & Model dependence      &  0--3     &  0--3     &  4--14  \\
  				  & {Total}                       &  {4}  &  {4}  & {5--14}  \\[\cmsTabSkip]
   \multirow{5}{*}{$D(\ptmin)$}
    				  & Tracking efficiency       &  4  &  4  & 4  					\\
  			 	  & Pileup modeling           &  0--2     &  0.5--2  &  0--2    \\
   				  & Event selection            &  $<$0.2  &  1           &  1--12  \\
   				  & Model dependence      &  0--4     &  0--4     &  1--11  \\
   				  & {Total}                       &  {4}  &  {4}  & {4--15}	\\ \hline
   \end{tabular}
     \label{table:systematics}
\end{table*}

\section{Results}
\label{sec:res}
\begin{figure*}[hbtp]
    \centering
    \includegraphics[width=0.9\cmsFigWidth]{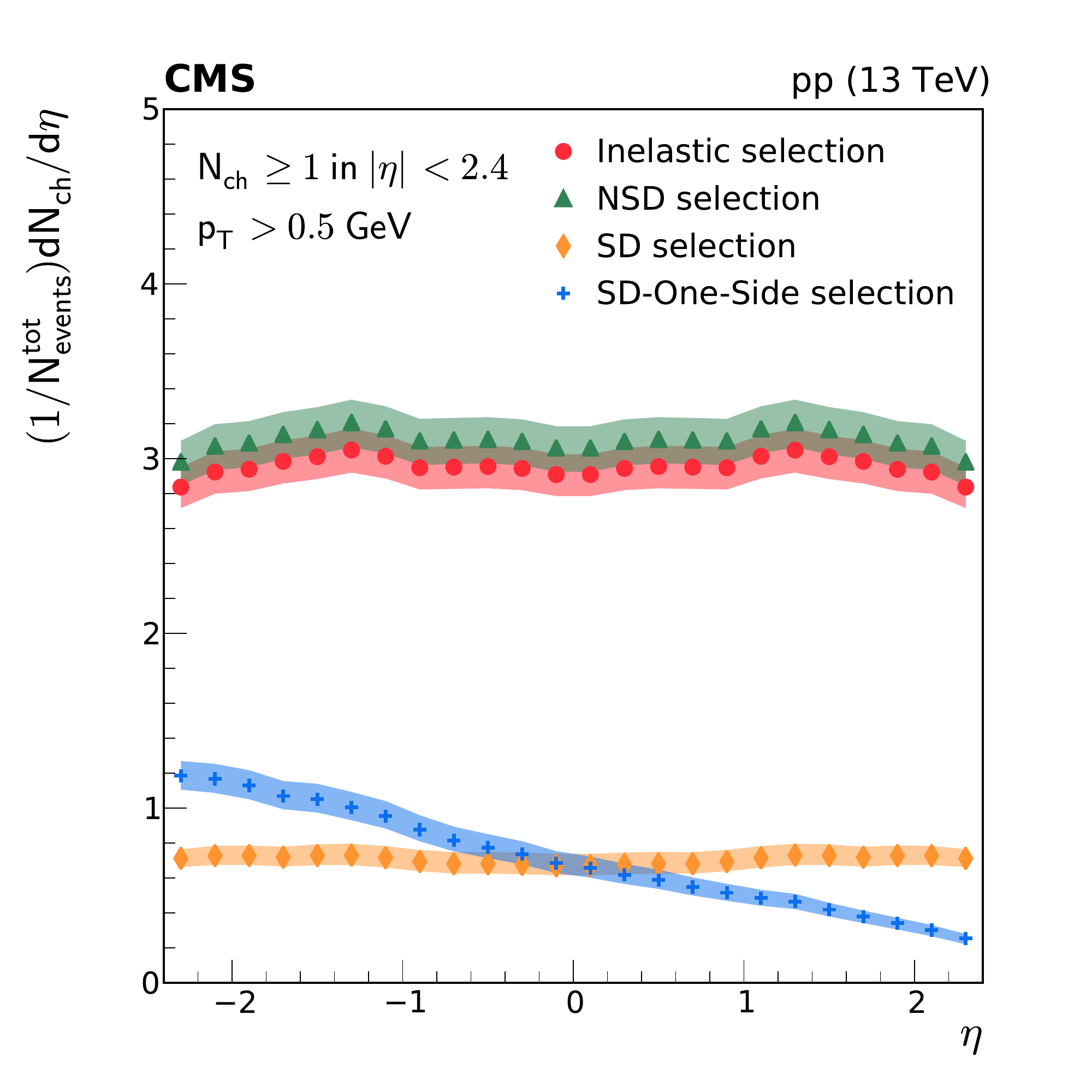}
    \includegraphics[width=0.9\cmsFigWidth]{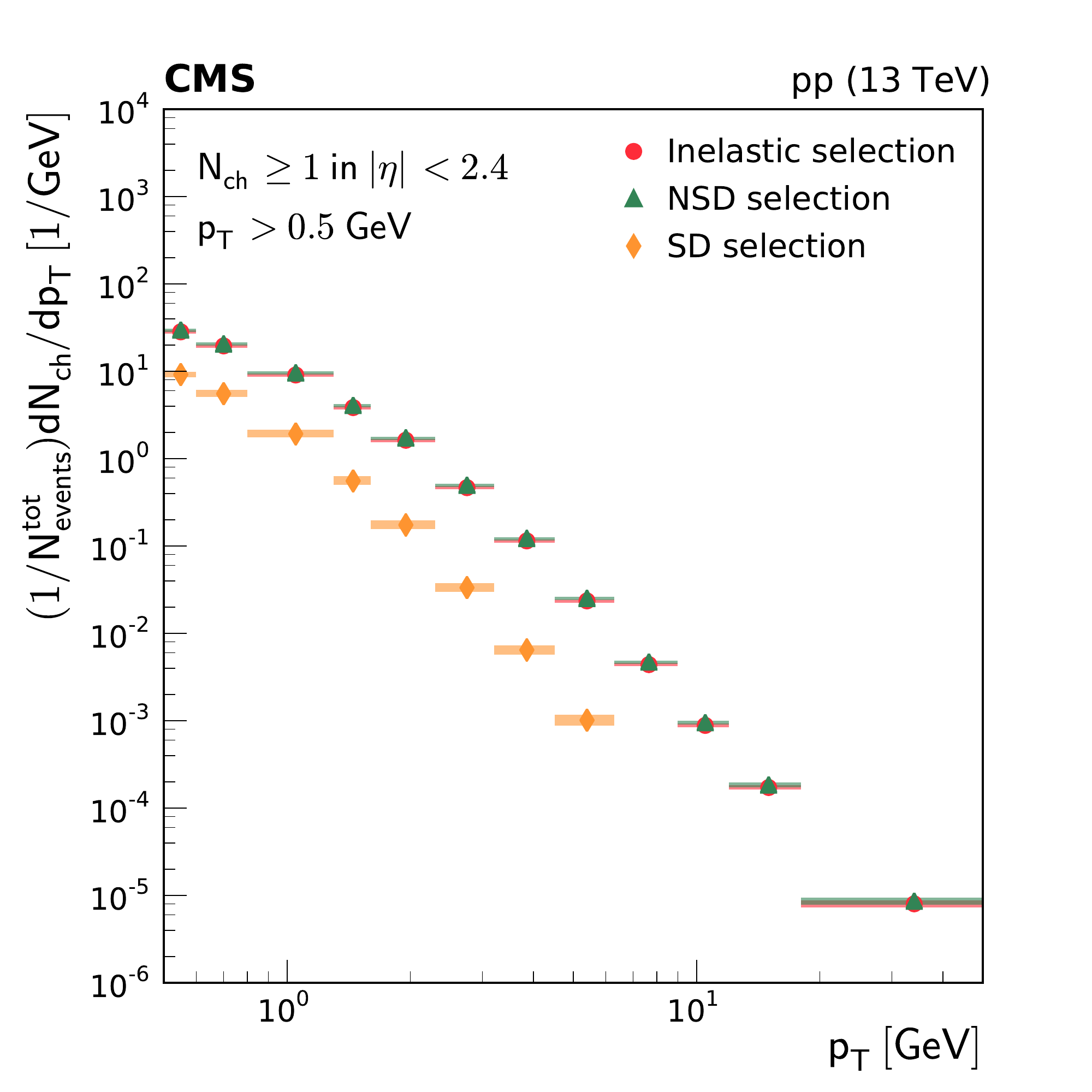}
    \includegraphics[width=0.9\cmsFigWidth]{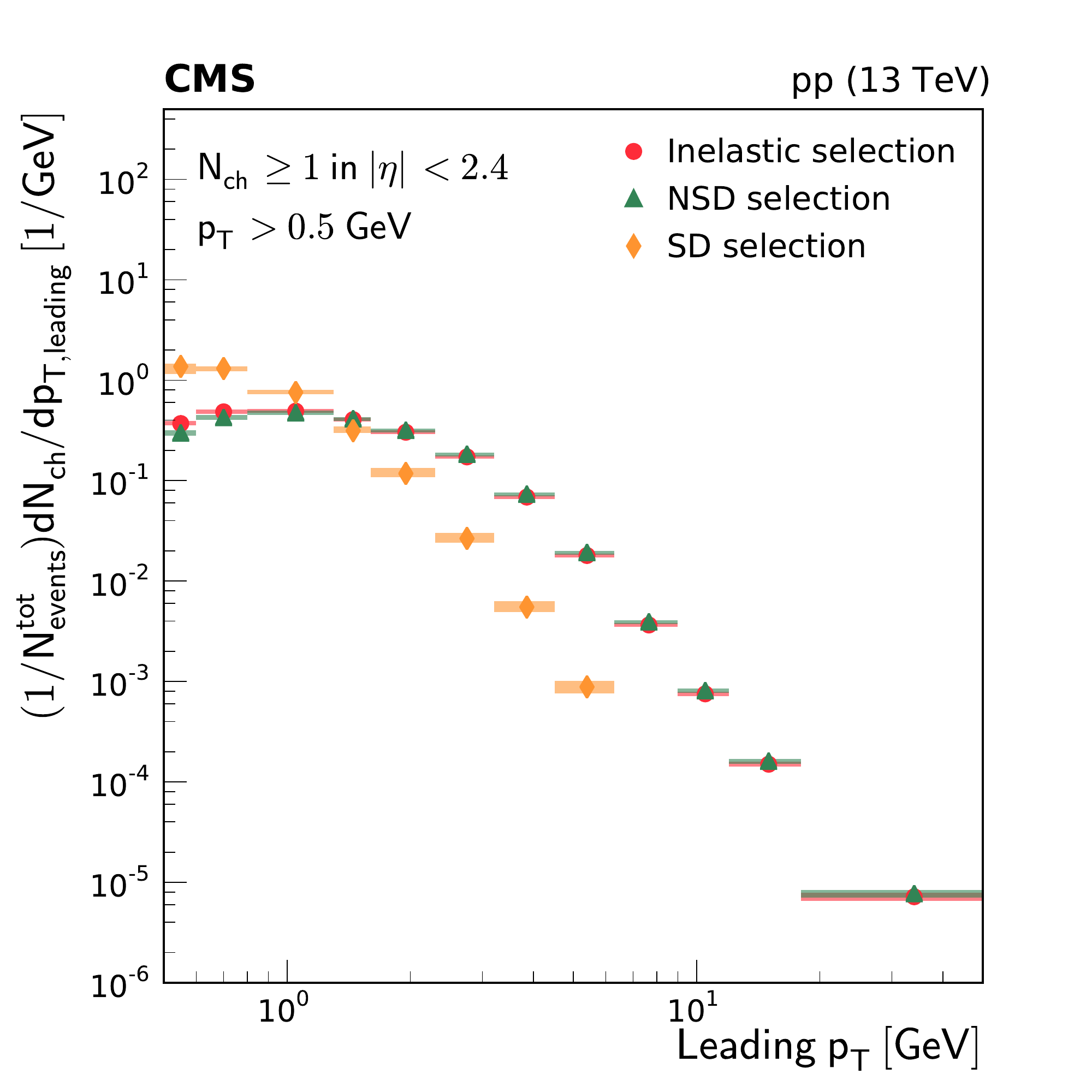}
    \includegraphics[width=0.9\cmsFigWidth]{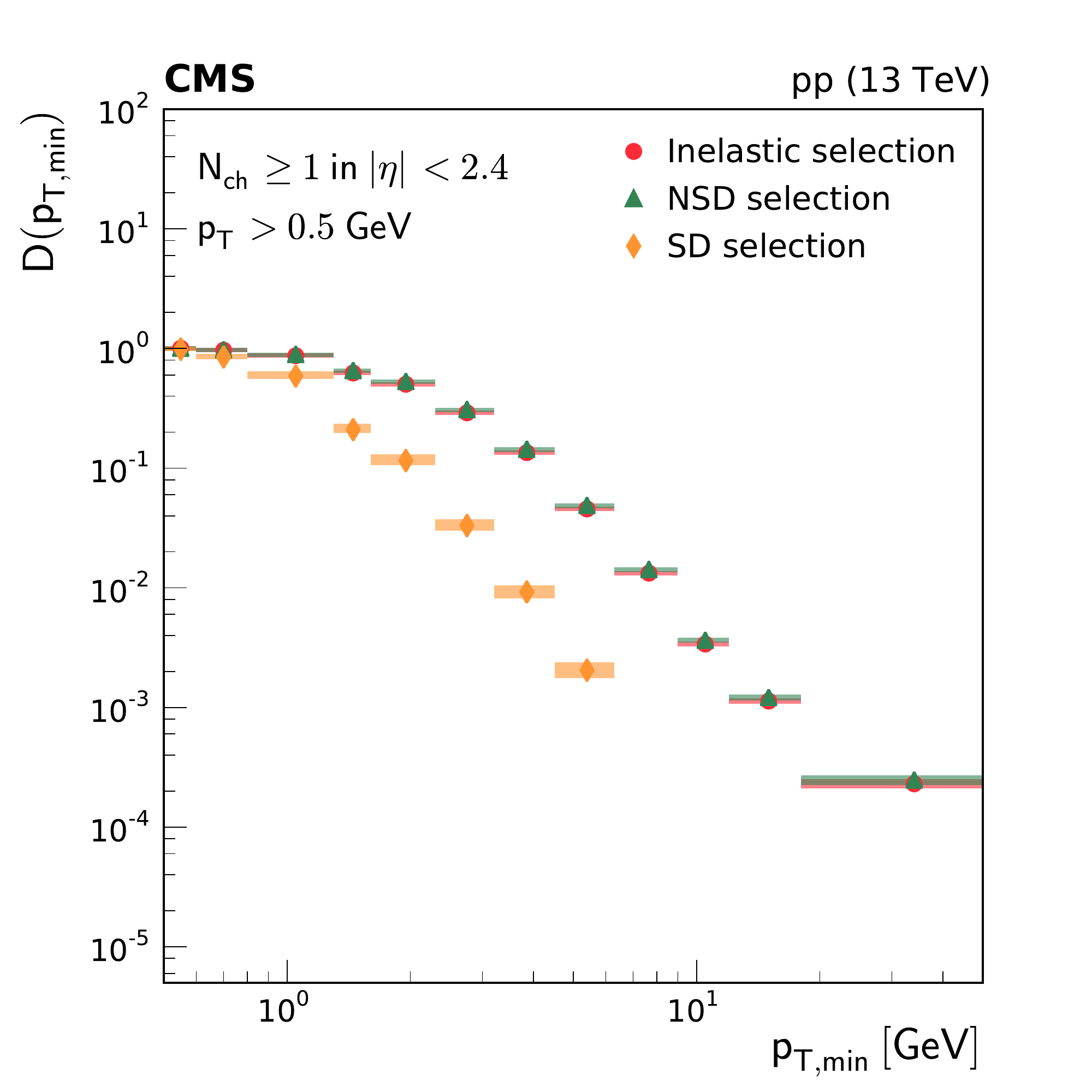}
    \includegraphics[width=0.9\cmsFigWidth]{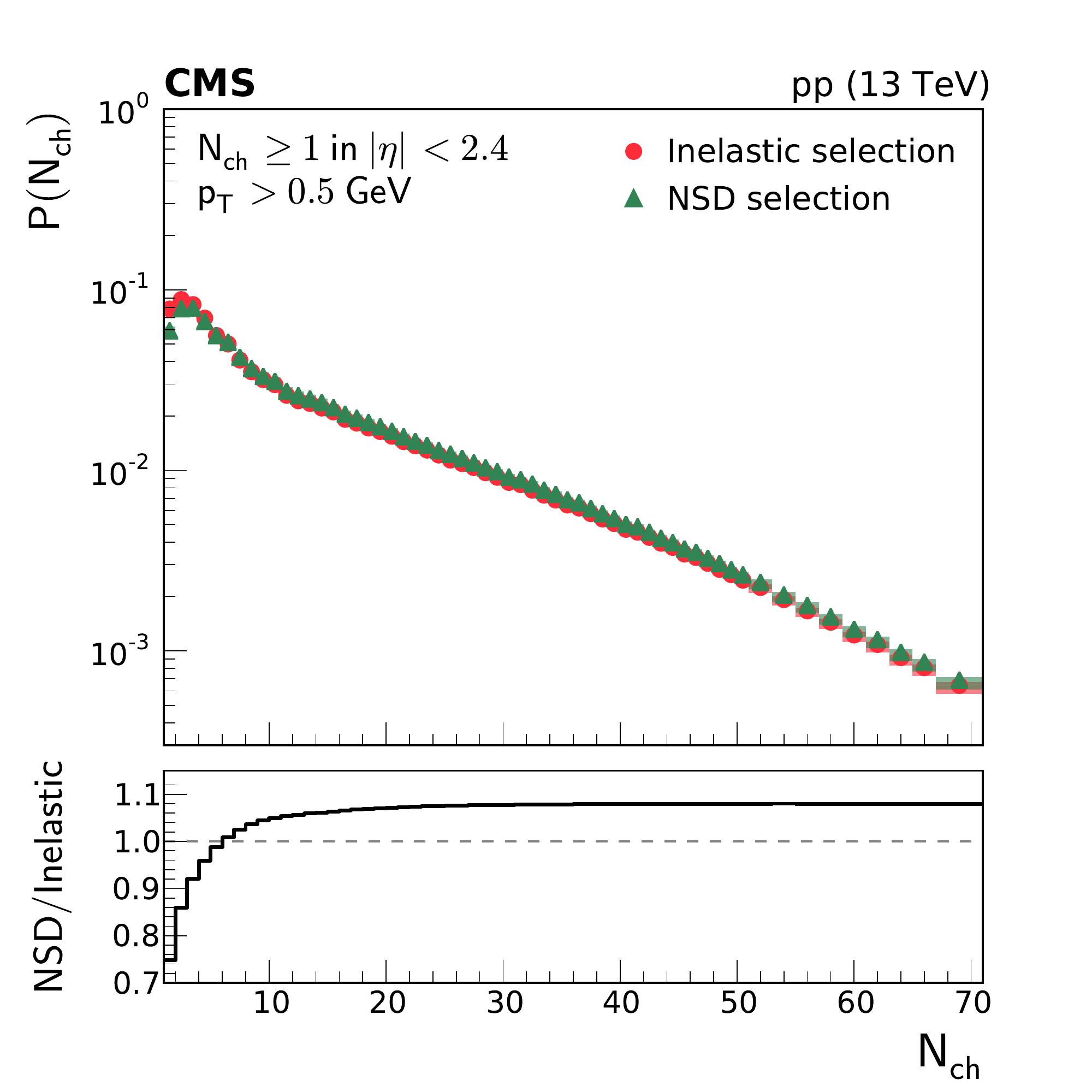}
      \caption{
     From top to bottom, left to right: \pse , \PT, leading \PT, integrated leading \PT, and multiplicity of charged particles per event  for the inelastic (circles), NSD-enhanced (triangles), SD-enhanced (diamonds), and SD-One-Side enhanced (crosses) event samples.
     The band encompassing the data points represent the total systematic uncertainty, while the statistical uncertainty is included as a vertical bar for each data point.
    }
    \label{fig:combined-results}
\end{figure*}

Charged particle distributions corrected to  particle level  as a function of  $\eta$, \PT and  leading \PT, as well as the integrated leading \PT  as a function of \ptmin  ($D(\ptmin)$), and the multiplicity per event $P(\nch)$ are shown in Fig.~\ref{fig:combined-results}.
They are presented for the different event categories corresponding to the most inclusive (inelastic), the diffraction-depleted (NSD), and the diffraction-enhanced (SD) samples.

The SD-minus and SD-plus samples are mutually exclusive, depending on the side of the forward-detector that contains the hadronic activity.
The \pse distribution of the SD-enhanced event sample is also presented as a symmetrized distribution constructed from the SD-minus and SD-plus enhanced samples and is referred to as the SD-One-Side enhanced event sample.
The symmetrization is performed by reflecting the distribution with respect to
$\eta = 0$.
The \pse distributions are averaged over the positive and negative \Eta ranges to suppress statistical fluctuations.

The per-event yields, defined in  Eq.~(\ref{int}), are obtained experimentally as
\begin{equation}
{D}(\ptmin)
= \frac {1}{N_{\text{events}}} \sum_{\ptlead>\ptmin} \hspace{-2.5em} \Delta \ptlead
\left(\dd{\Delta N}{\Delta \ptlead}\right),
\end{equation}
 \begin{tolerant}{800}
where $N_{\text{events}}$ is the number of events with a leading charged particle, $\Delta \ptlead$ is the bin width, and $\Delta N$ is the number of events with a leading charged particle in each bin.
\end{tolerant}

 \begin{tolerant}{800}
In general terms, the inelastic and NSD distributions are similar.  The \pse density of the SD-enhanced event sample is about a factor of 4
lower than that of the most inclusive event samples.
The \PT distributions (\ie,~\PT, leading \PT, and integrated leading \PT) of the SD-enhanced event sample fall very  steeply  for large \pt values.
The charged particle multiplicity distribution of  the NSD-enhanced event sample shows a depletion of low-multiplicity events and an increase of high-multiplicity events compared to that of the inelastic sample.
\end{tolerant}

\begin{figure*}[hbtp]
    \centering
    \includegraphics[width=\cmsFigWidth]{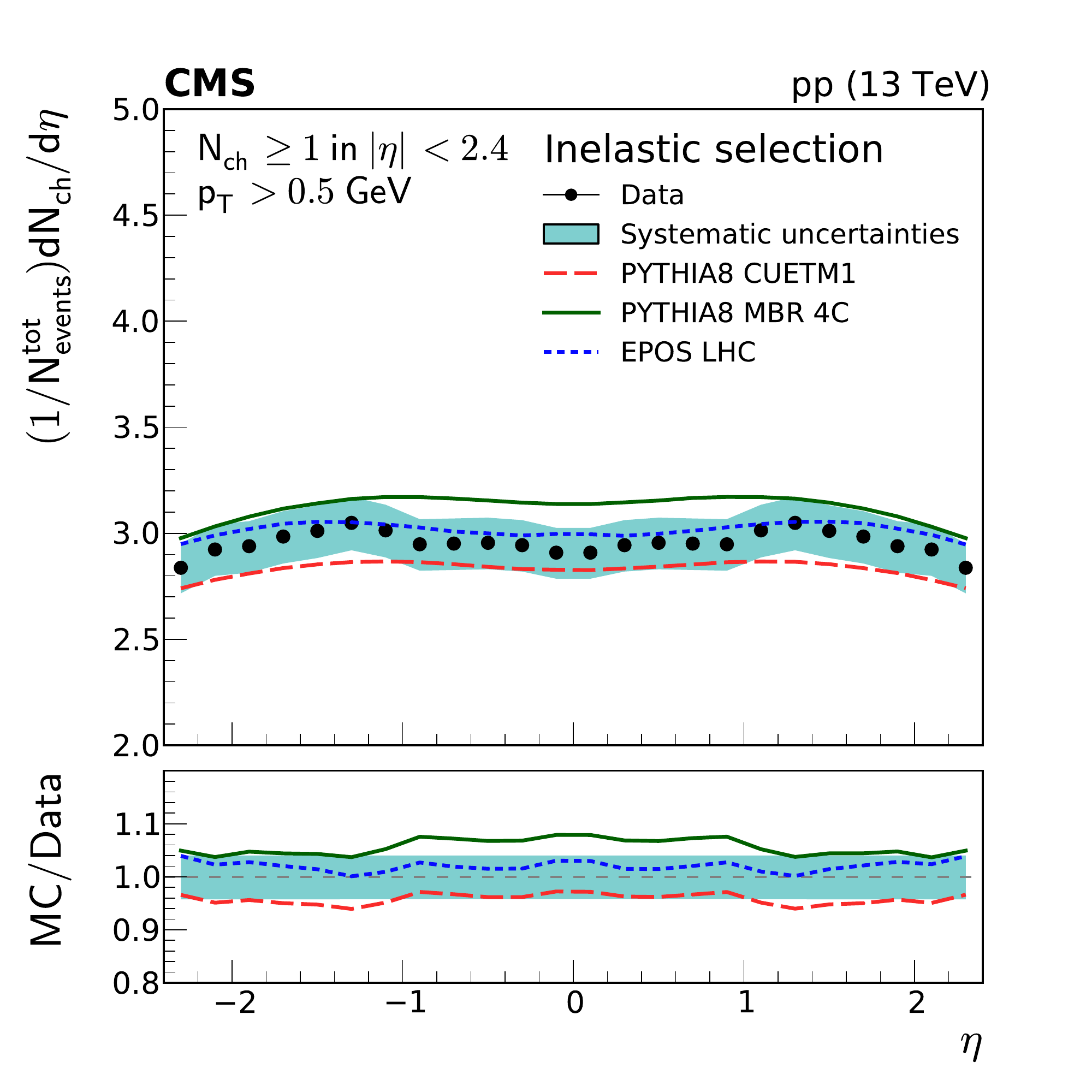}
    \includegraphics[width=\cmsFigWidth]{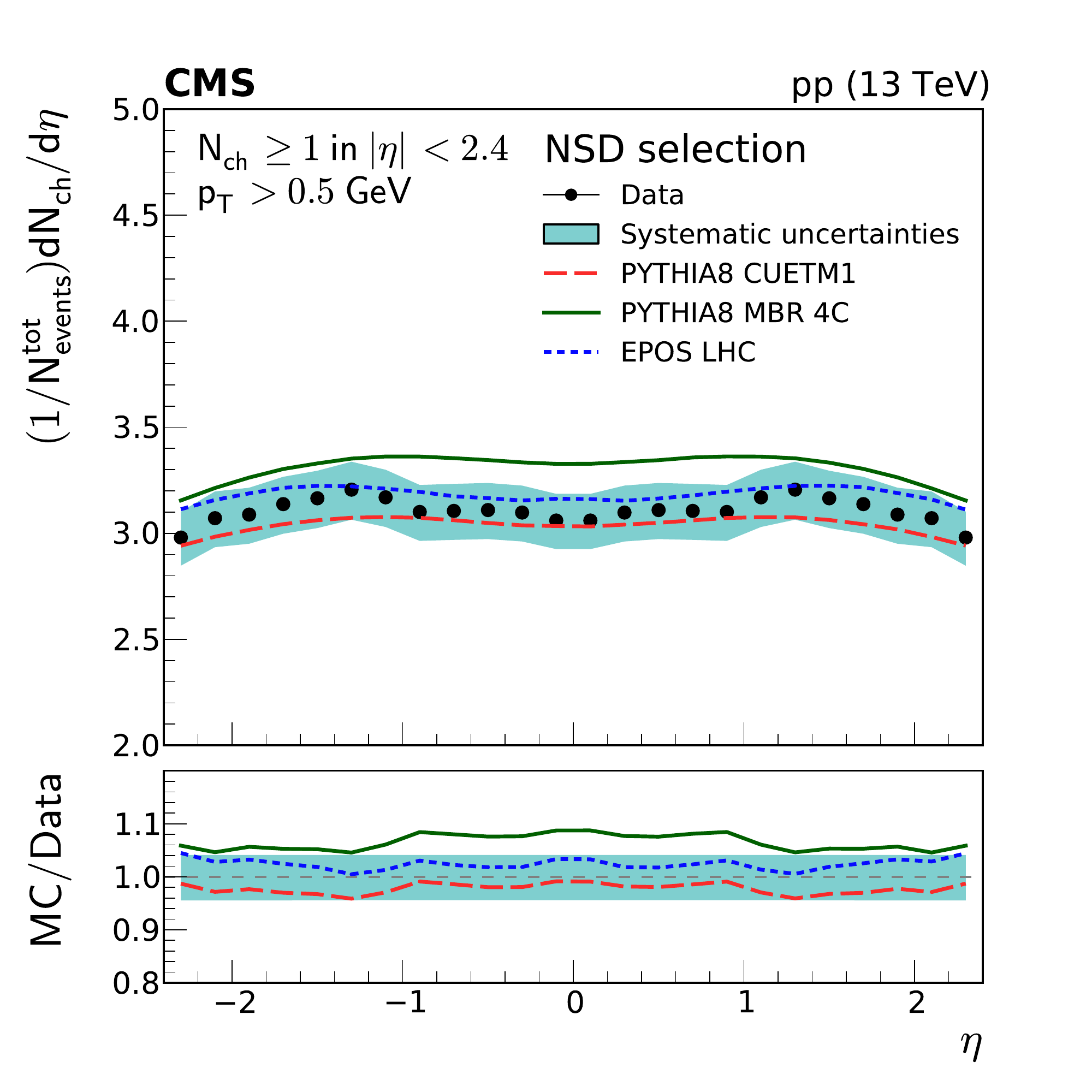}
    \includegraphics[width=\cmsFigWidth]{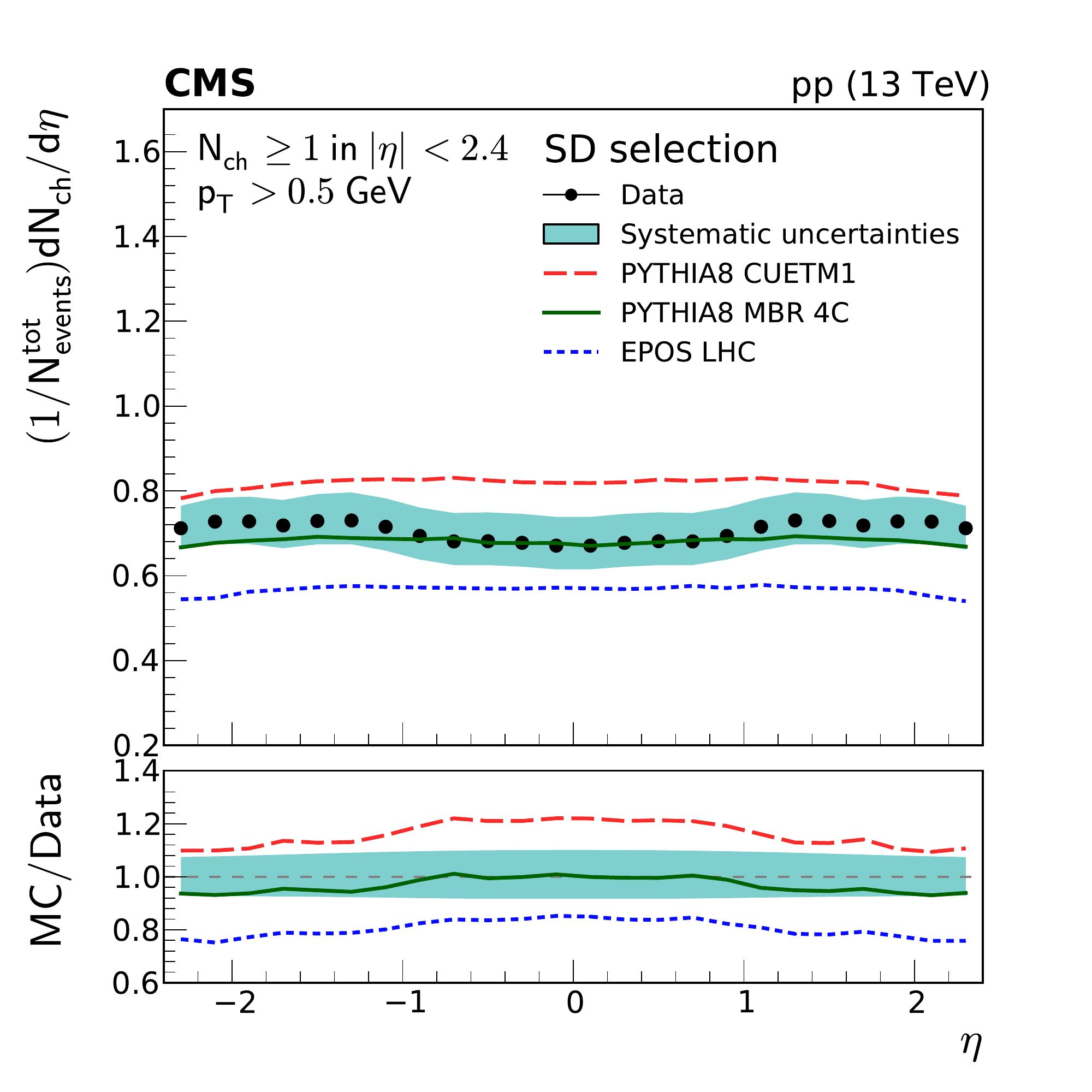}
    \includegraphics[width=\cmsFigWidth]{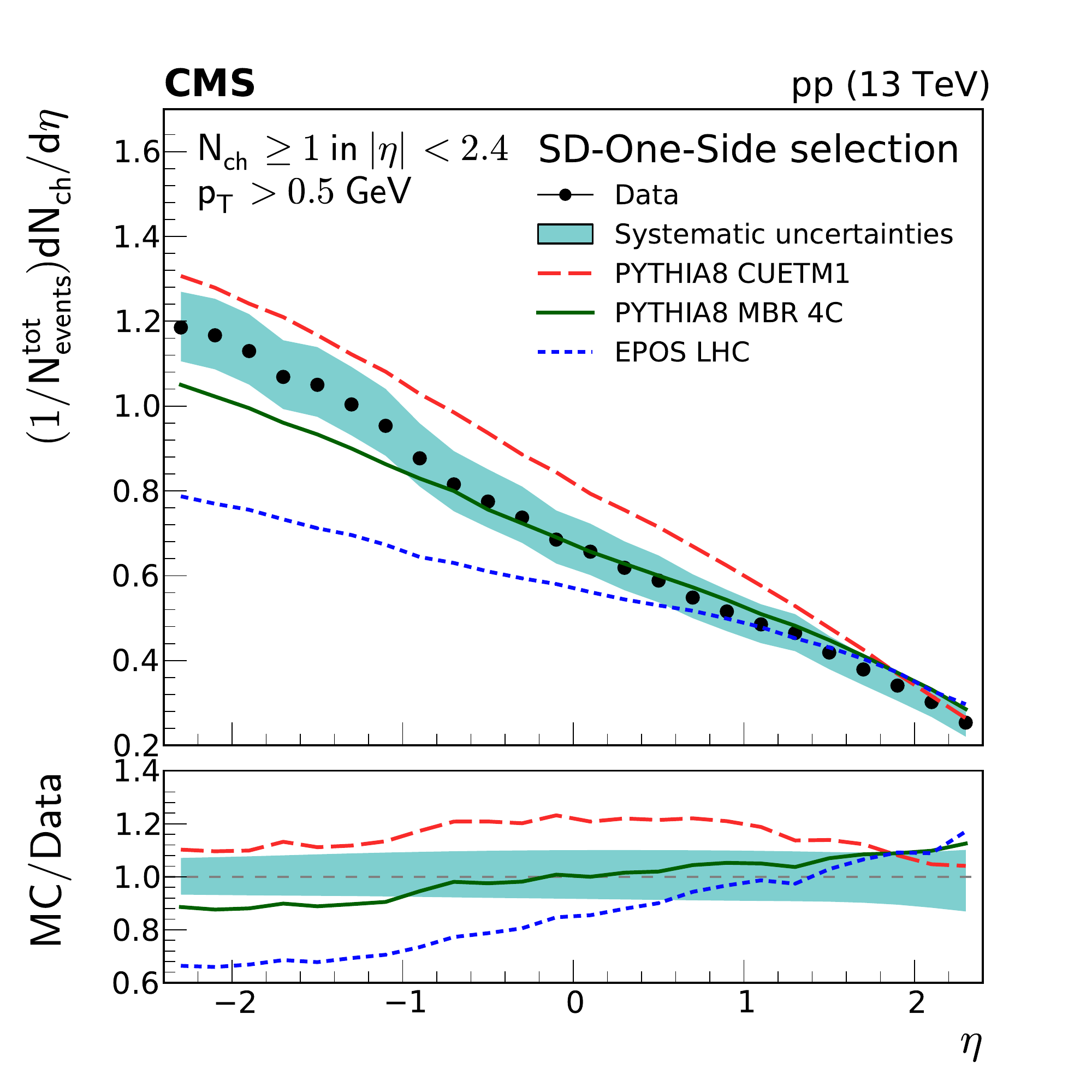}
  \caption{
Charged particle \pse densities averaged over both positive and negative \Eta ranges.
Top to bottom, left to right: inelastic, NSD-, SD-, and SD-One-Side enhanced event samples.
The measurements are compared to the predictions of the  \PYTHIA{}8 \CUETM (long dashes), \PYTHIA{}8 \MBR (continuous line), and \EPOS\ \LHC (short dashes) event generators.
The band encompassing the data points represent the total systematic uncertainty, while the statistical uncertainty is included as a vertical bar for each data point.
The lower panels show the corresponding MC-to-data ratios.}
    \label{fig:result-eta}
\end{figure*}

 \begin{tolerant}{800}
Figure~\ref{fig:result-eta} shows the \pse densities of charged particles for four different event categories.
The measurements are compared to the predictions of different MC event generators, namely \PYTHIA{}8 \CUETM, \PYTHIA{}8 \MBR , and \EPOS\ \LHC.
The predictions of \EPOS\ \LHC provide the best description of the data within uncertainties for the inelastic event sample.
The predictions of \PYTHIA{}8 \CUETM slightly underestimate the measurements, while those of \PYTHIA{}8 \MBR overestimate them.
For the NSD-enhanced event sample, the predictions of \EPOS\ \LHC and \PYTHIA{}8 \CUETM both give a reasonable description of the data within uncertainties, while the predictions of \PYTHIA{}8 \MBR overestimate the measurements.
The opposite behavior is observed for the SD-enhanced event samples, with the predictions of \EPOS\ \LHC underestimating the data, and the predictions of \PYTHIA{}8 \CUETM overestimating them.
The predictions from \PYTHIA{}8 \MBR describe well the SD data within uncertainties, showing only small deviations at the edges of the phase space.
The SD-One-Side enhanced event sample is not well described by \EPOS\ \LHC,  while \PYTHIA{}8 \CUETM tends to overestimate data,  and the prediction from  \PYTHIA{}8 \MBR describes the measurements within uncertainties over almost the full range, exhibiting some deviations in the regions where the diffractive dissociative system is observed.
\end{tolerant}

\begin{figure*}[htbp]
    \centering
    \includegraphics[width=\cmsFigWidth]{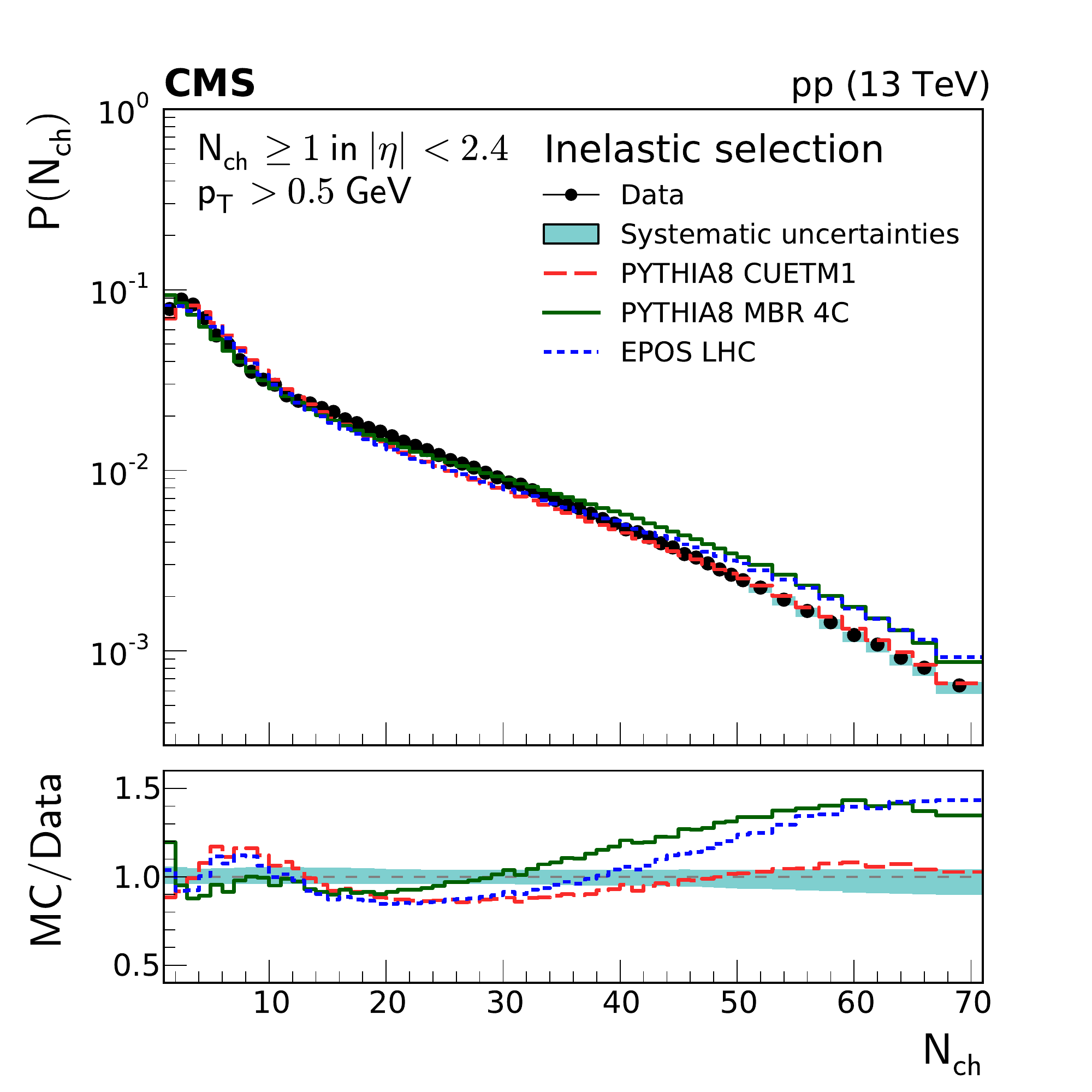}
    \includegraphics[width=\cmsFigWidth]{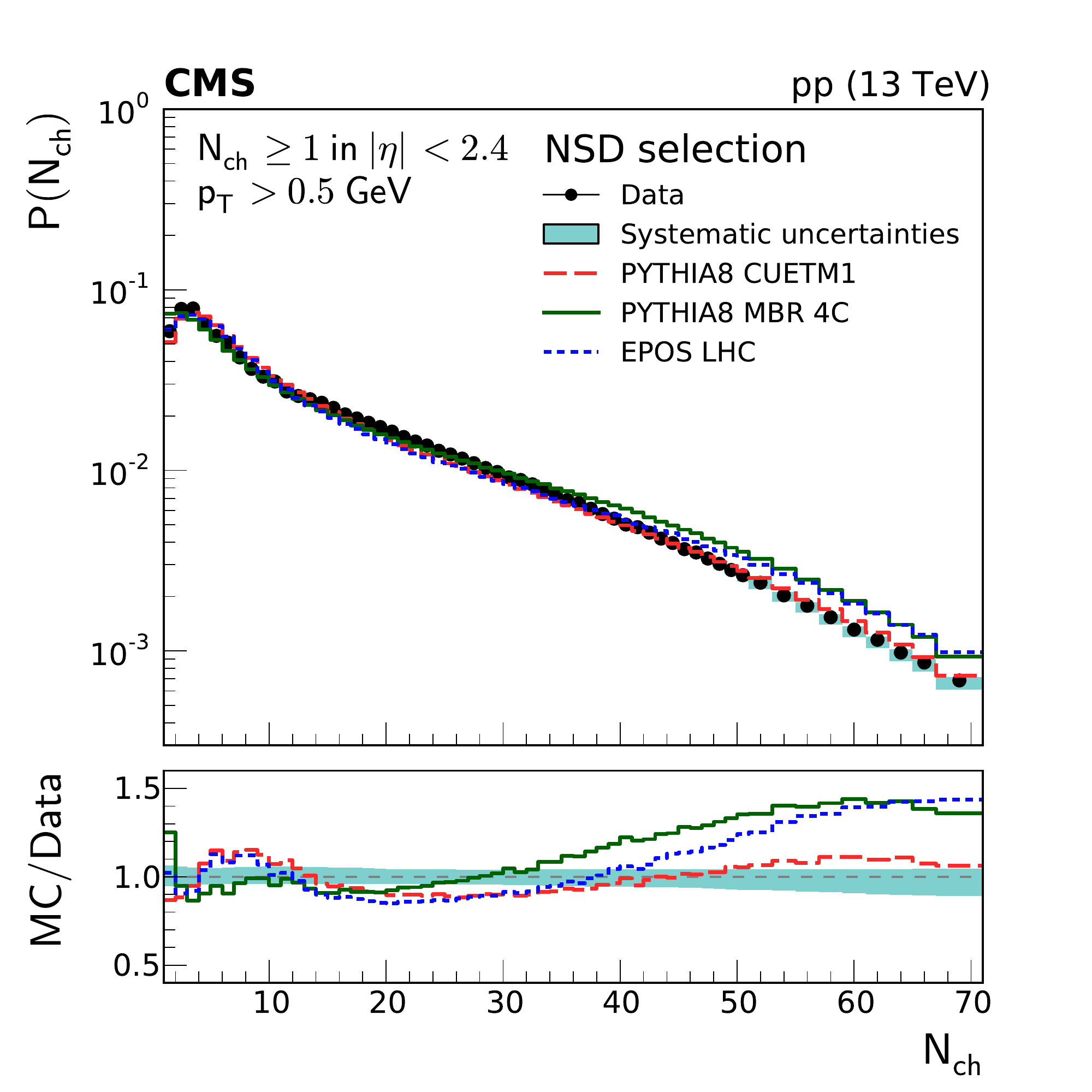}
   \caption{
Charged particle multiplicity distributions of the  inelastic (left), and NSD-enhanced (right) event samples.
The measurements are compared to the predictions of the  \PYTHIA{}8 \CUETM (long dashes), \PYTHIA{}8 \MBR (continuous line), and \EPOS\ \LHC (short dashes) event generators.
The band encompassing the data points represent the total systematic uncertainty, while the statistical uncertainty is included as a vertical bar for each data point.
The lower panels show the corresponding MC-to-data ratios.}
    \label{fig:result-M}
\end{figure*}

Figure~\ref{fig:result-M} shows the charged particle multiplicity distributions for the inelastic and NSD-enhanced event samples.
The different event generators provide similar predictions for the inelastic and NSD-enhanced event samples, with  differences appearing only at low multiplicities.
It is in this  low-multiplicity regime that the SD dissociation events contribute the most.
The \PYTHIA{}8 \MBR generator gives the best description of the data in the low-multiplicity region, while \PYTHIA{}8 \CUETM and \EPOS\ \LHC overestimate the data by approximately 20\%.
This behavior is similar to that observed in the \pse distribution of the SD-enhanced selection, where \PYTHIA{}8 \MBR provides the best description of  the SD-enhanced event sample.
For multiplicities above~35, the predictions of \PYTHIA{}8 \CUETM give the best description of the data, whereas those of \PYTHIA{}8 \MBR  and \EPOS\ \LHC are off by up to 50\%.
The high multiplicity region is especially sensitive to MPI and improving its modelling could lead to a better understanding of these processes.

\begin{figure*}[htbp]
    \centering
    \includegraphics[width=\cmsFigWidth]{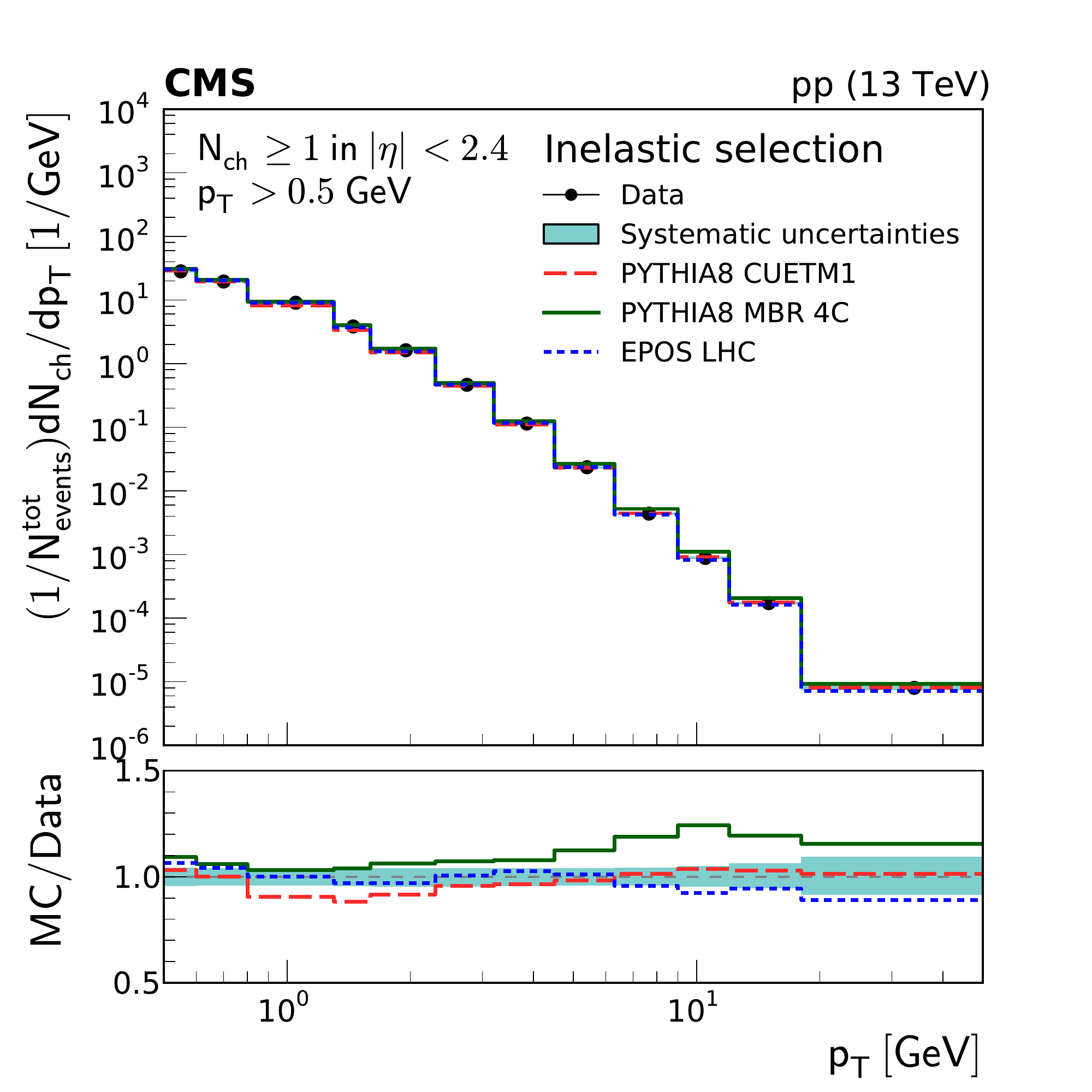}
    \includegraphics[width=\cmsFigWidth]{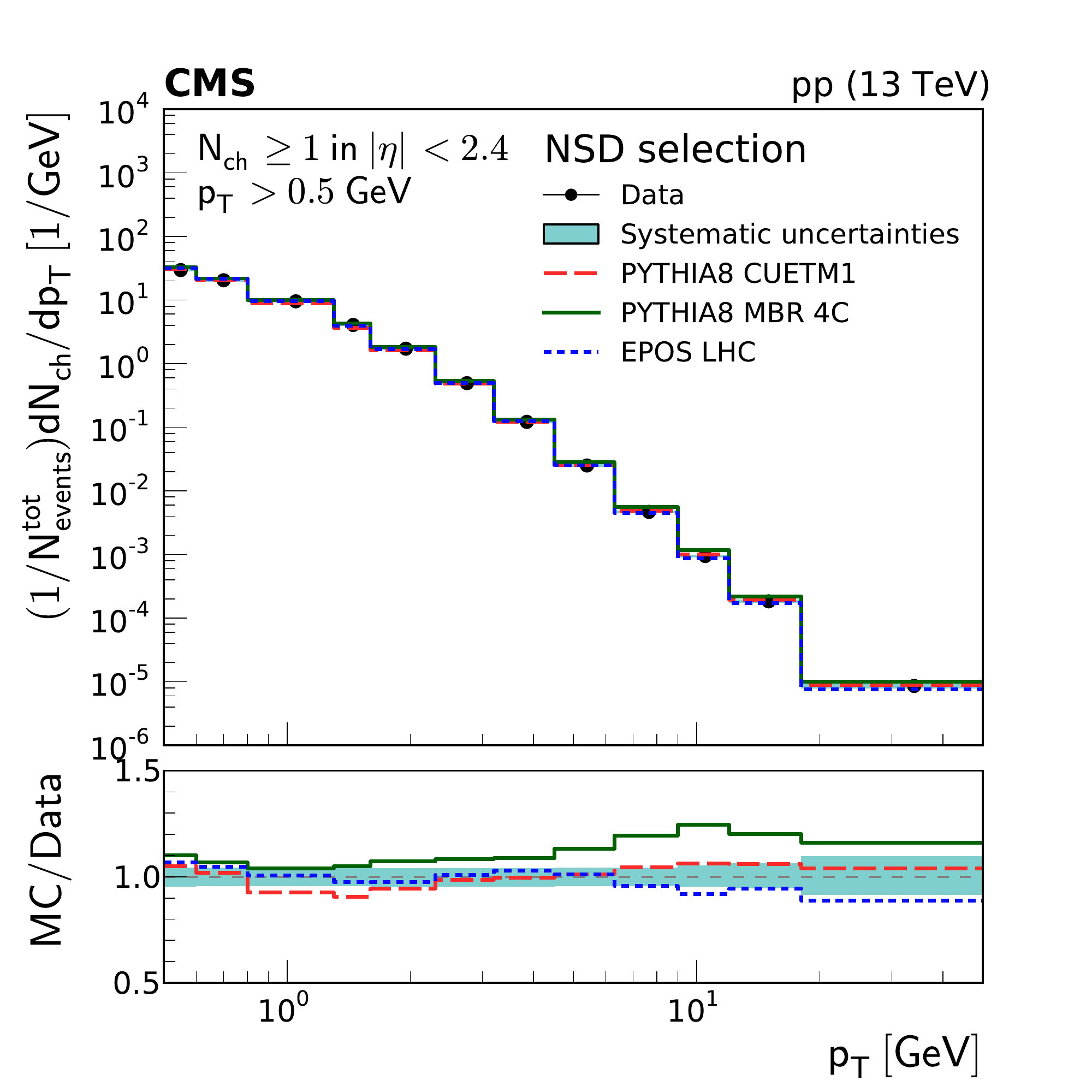}
    \includegraphics[width=\cmsFigWidth]{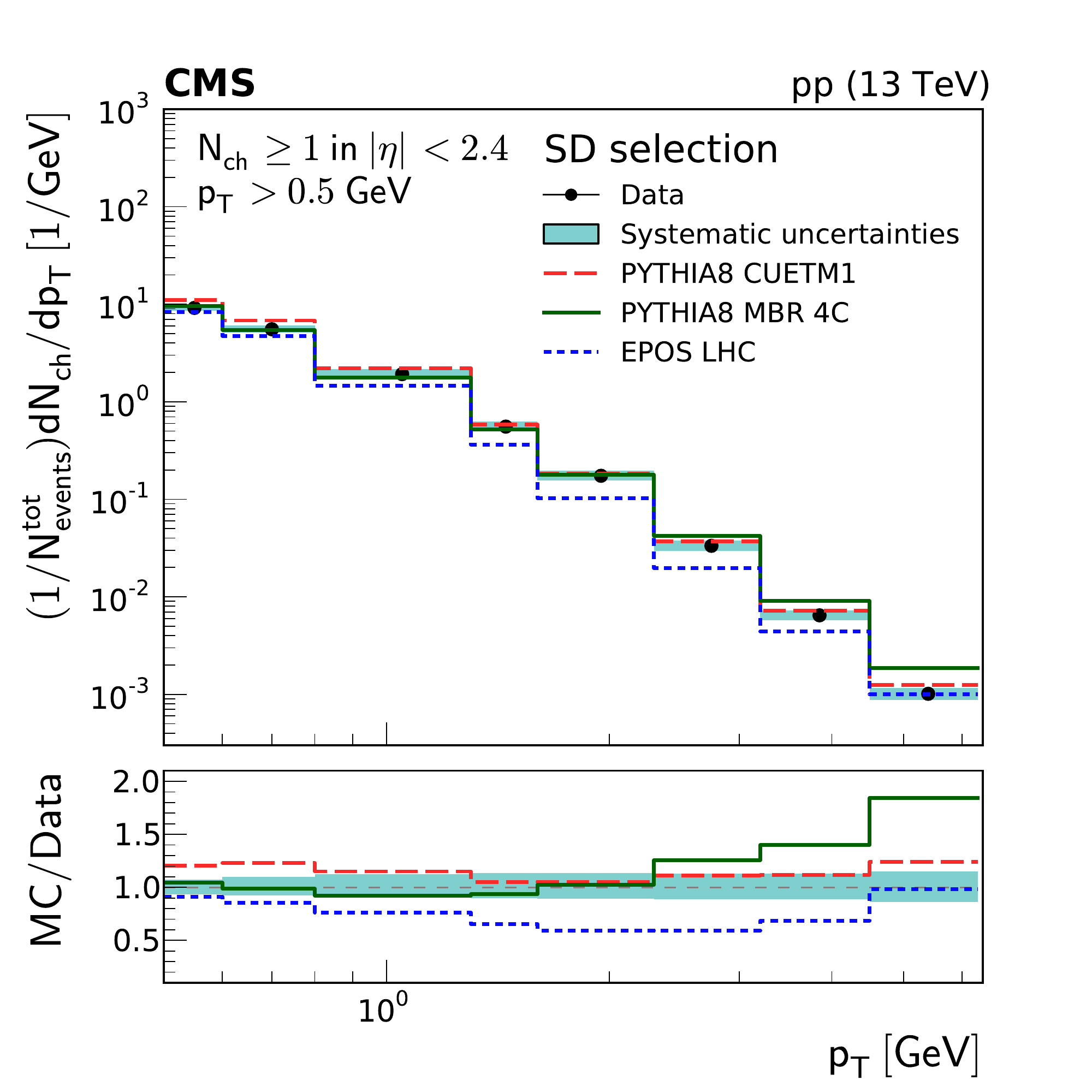}
   \caption{
Charged particle transverse-momentum densities of inelastic (top left), NSD-enhanced (top right), and SD-enhanced (bottom) event samples.
The measurements are compared to the predictions of the  \PYTHIA{}8 \CUETM (long dashes), \PYTHIA{}8 \MBR (continuous line), and \EPOS\ \LHC (short dashes) event generators.
The band encompassing the data points represent the total systematic uncertainty, while the statistical uncertainty is included as a vertical bar for each data point.
The lower panels show the corresponding MC-to-data ratios.}
    \label{fig:result-pt}
\end{figure*}

\begin{figure*}[htbp]
    \centering
    \includegraphics[width=\cmsFigWidth]{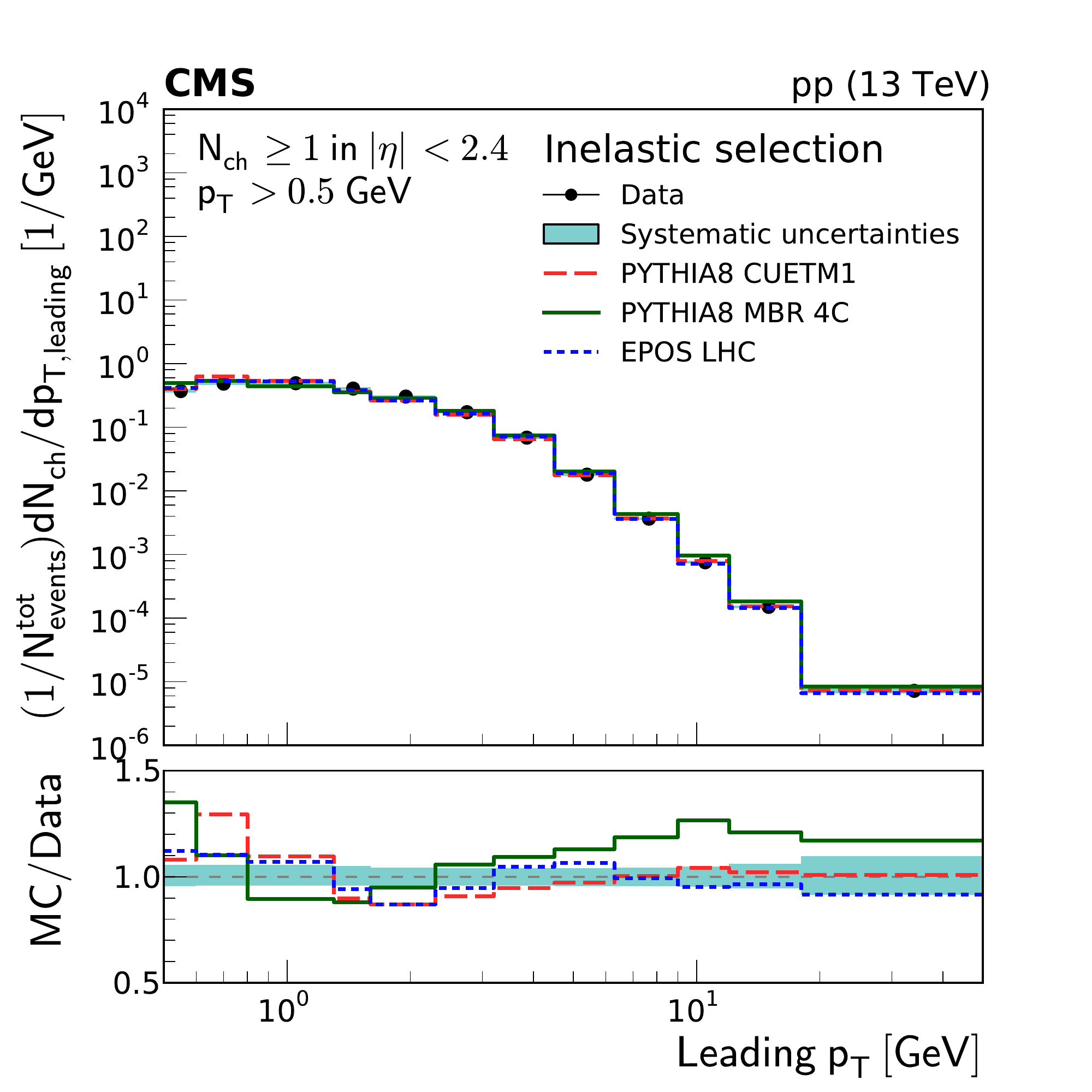}
    \includegraphics[width=\cmsFigWidth]{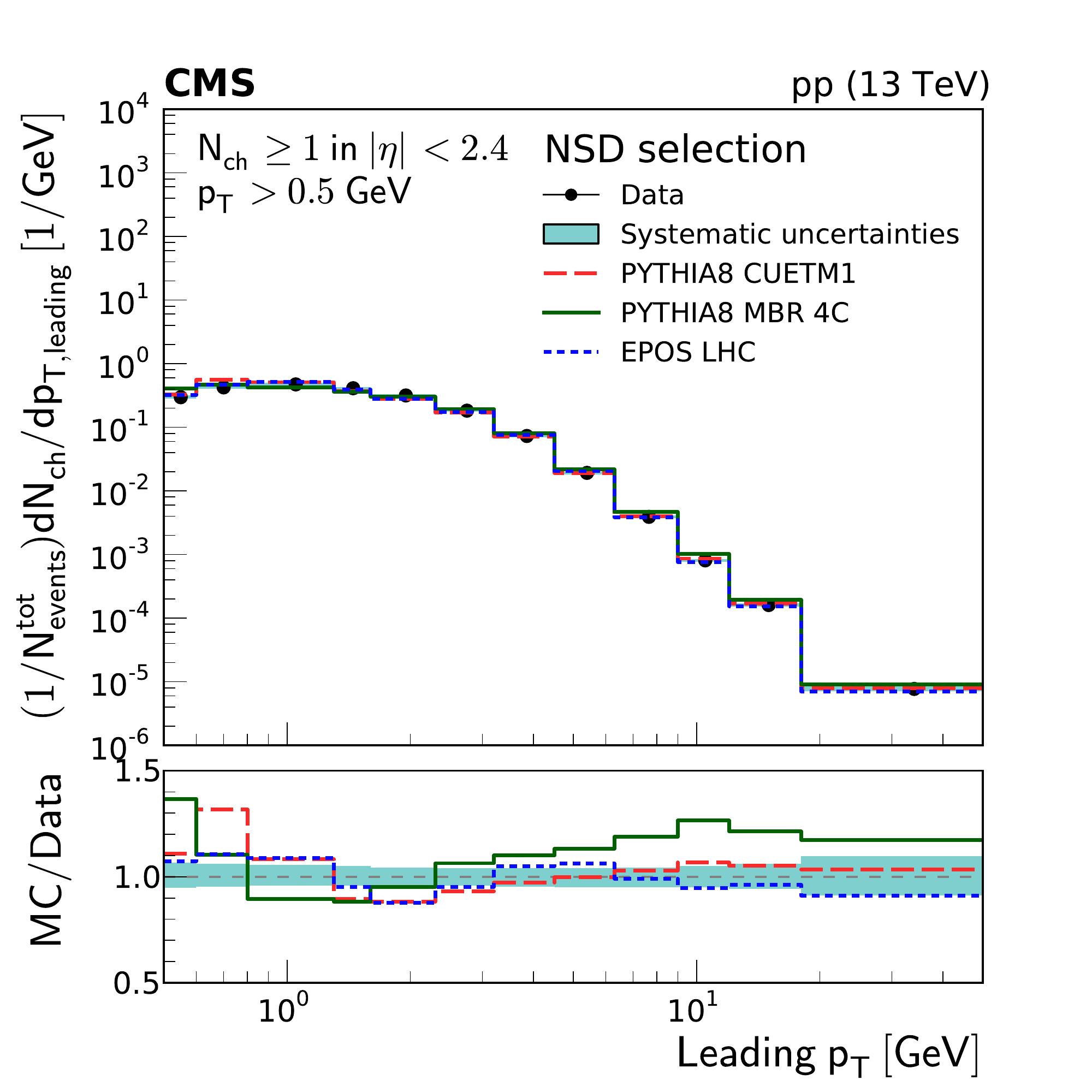}
    \includegraphics[width=\cmsFigWidth]{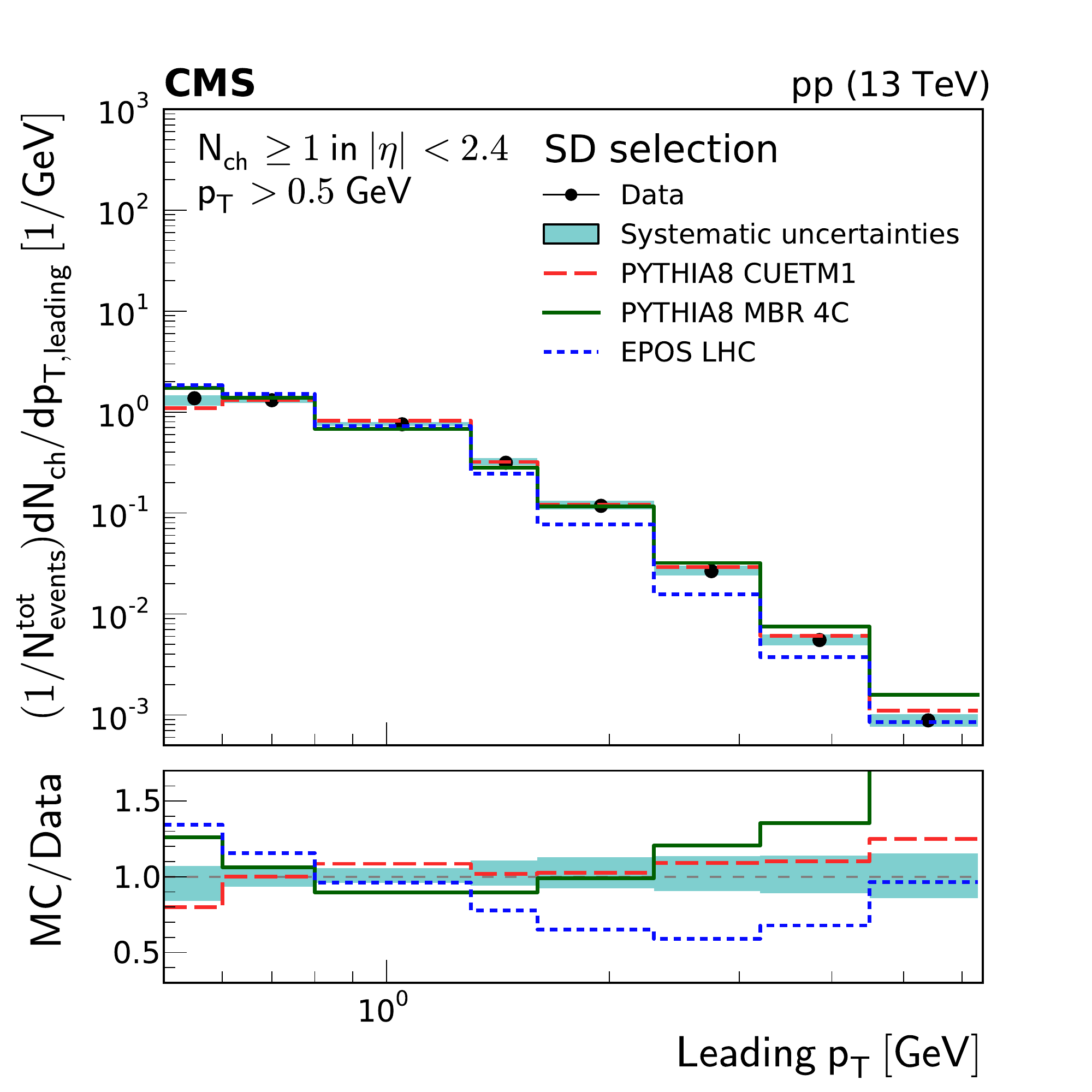}
    \caption{
Leading charged particle \PT distributions of inelastic (top left), NSD-enhanced (top right), and SD-enhanced  (bottom) event samples.
The measurements are compared to the predictions of the  \PYTHIA{}8 \CUETM (long dashes), \PYTHIA{}8 \MBR (continuous line), and \EPOS\ \LHC (short dashes) event generators.
The band encompassing the data points represent the total systematic uncertainty, while the statistical uncertainty is included as a vertical bar for each data point.
The lower panels show the corresponding MC-to-data ratios.}
    \label{fig:result-pt-leading}
\end{figure*}

\begin{figure*}[hbtp]
    \centering
    \includegraphics[width=\cmsFigWidth]{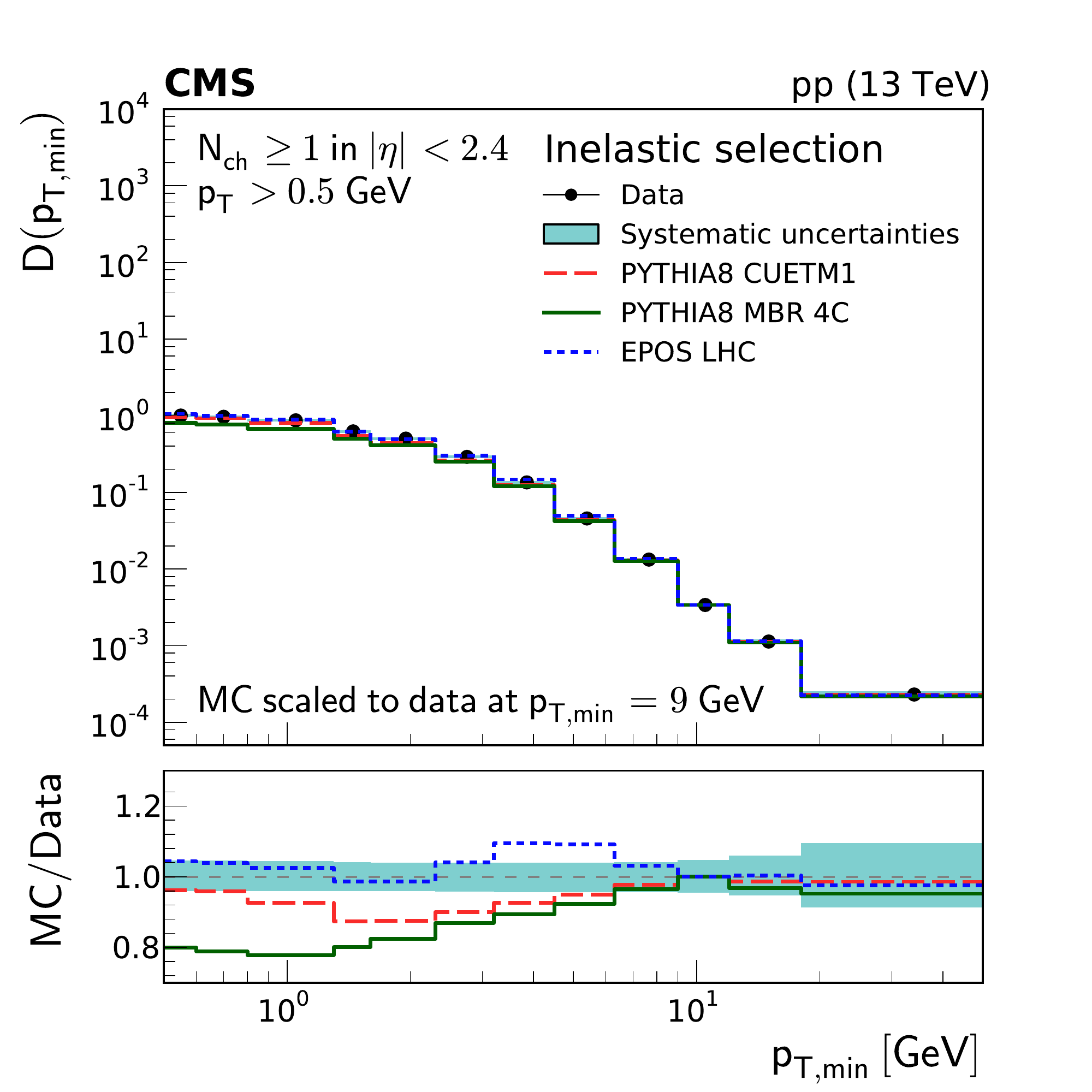}
    \includegraphics[width=\cmsFigWidth]{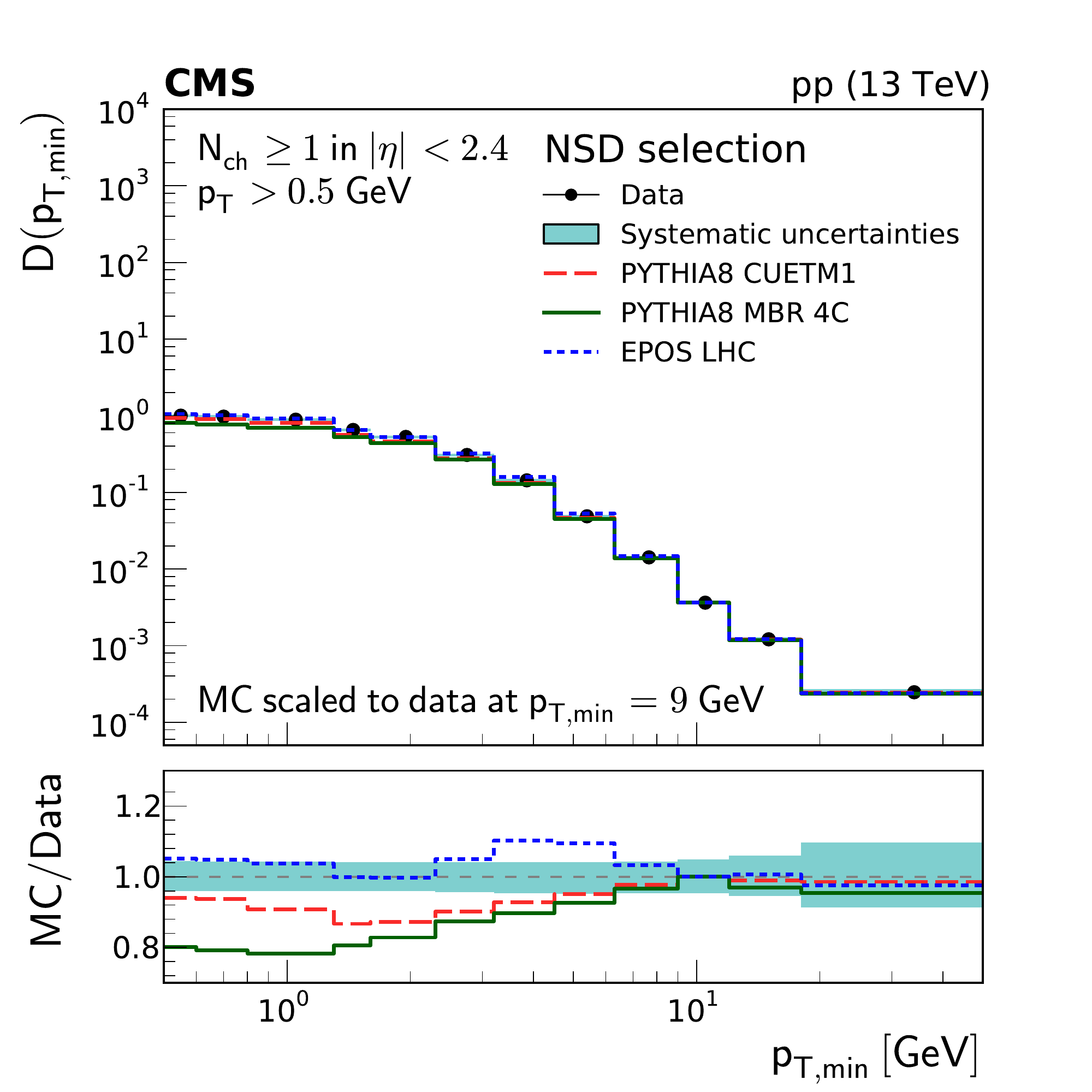}
    \includegraphics[width=\cmsFigWidth]{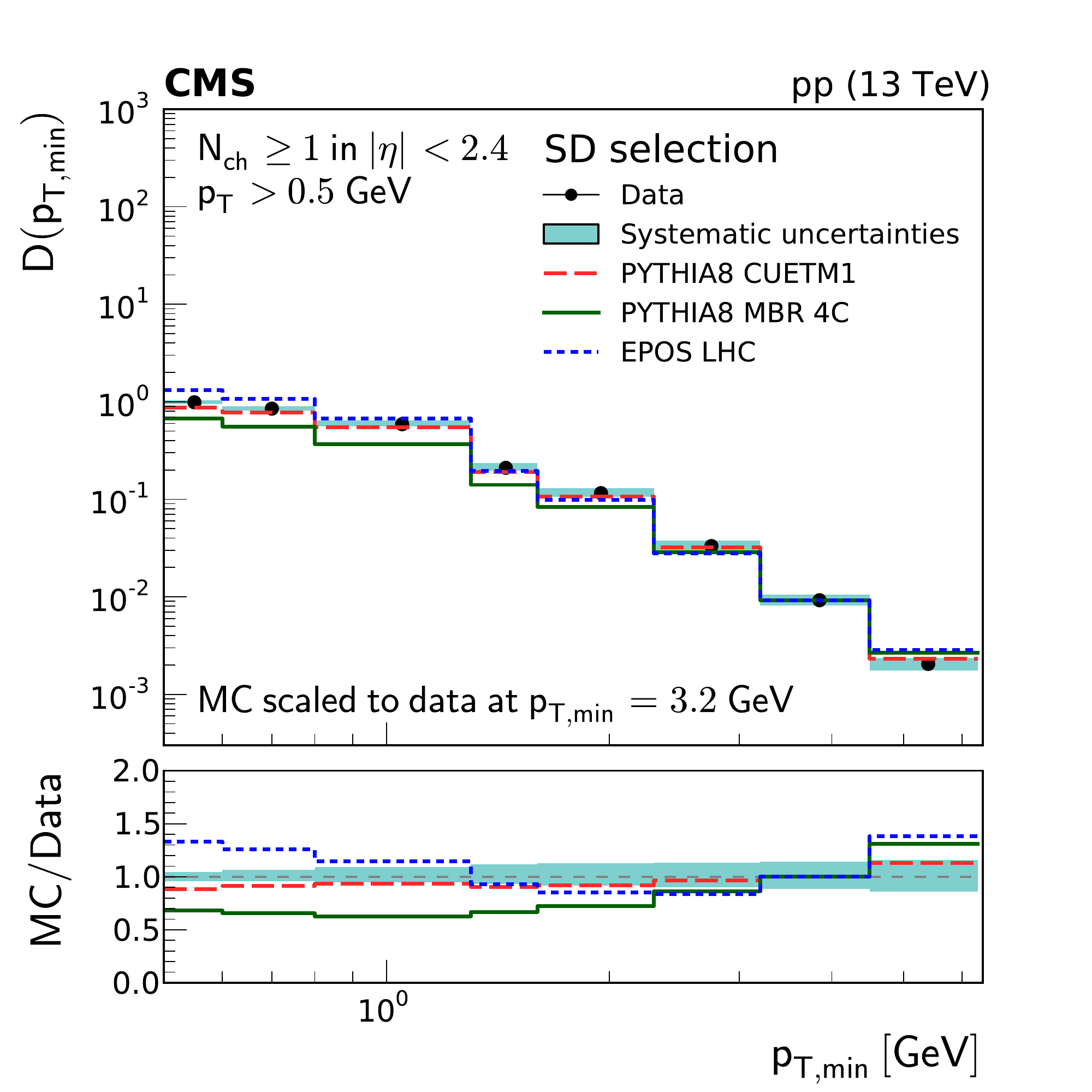}
   \caption{
Integrated leading charged particle \PT distributions as a function of \ptmin for
inelastic (top left), NSD-enhanced (top right), and SD-enhanced  (bottom) event samples.
The measurements are compared to the predictions of the  \PYTHIA{}8 \CUETM (long dashes), \PYTHIA{}8 \MBR (continuous line), and \EPOS\ \LHC (short dashes) event generators.
The band encompassing the data points represent the total systematic uncertainty, while the statistical uncertainty is included as a vertical bar for each data point.
The lower panels show the corresponding MC-to-data ratios.}
    \label{fig:result-pt-leading-integrated}
\end{figure*}

Figures~\ref{fig:result-pt} to \ref{fig:result-pt-leading-integrated} show the charged particle \PT distributions for all the particles, the leading particle, and the integrated spectrum of the latter, for the inelastic, NSD-, and SD-enhanced event samples.
The \pt range for the SD-enhanced event sample is smaller compared to the other samples, ranging up to 6.3\GeV instead of 50\GeV.
This is a consequence of the more steeply falling \pt spectrum of the SD-enhanced event sample with respect to the other two event categories (Fig.~\ref{fig:combined-results}).

The \PT distributions of the charged particles in the inelastic and NSD-enhanced event samples are best described by the predictions of \PYTHIA{}8 \CUETM over almost the full \pt range.
Small deviations of up to 10\% in the low-\pt region are observed.
This region is dominated by particles coming from MPI.
The predictions of \PYTHIA{}8 \MBR describe the low-\pt region but rapidly start to overestimate particle production for $\pt>5\GeV$
by up to 30\%.
The predictions of \EPOS\ \LHC give a reasonable description of the data for transverse momenta up to $\pt \approx10$\GeV, while above this value they underestimate it by $\approx10$\%.
In the case of the SD-enhanced event sample, \PYTHIA{}8 \CUETM gives the best description of the data, while \EPOS\ \LHC and \PYTHIA{}8 \MBR underestimate or overestimate the data by 40 and 80\%, respectively.
It is interesting to observe how difficult
it is to simultaneously describe within a given model both the bulk of soft particles mainly coming from MPI and the high-\pt particles primarily coming from hard parton scattering.

The leading \PT distributions of charged particles and their integral as a function of \ptmin are presented in Figs.~\ref{fig:result-pt-leading} and \ref{fig:result-pt-leading-integrated}, respectively.
These two distributions provide valuable information on the modeling of the transition between the nonperturbative and perturbative regimes, and on the modeling of MPI~\cite{Grebenyuk:2012qp}.
For the case of the leading transverse momentum distributions, the predictions of \EPOS\ \LHC give the best description of the data for the inelastic and NSD-enhanced event samples almost everywhere within the experimental uncertainties with only some small deviations at low-\pt values of up to $\approx$10\%.
For \mbox{$\pt > 4 \GeV$}, the predictions of \PYTHIA{}8 \CUETM are able to reproduce these data.
The predictions of \PYTHIA{}8 \MBR are not able to describe the data at either low or high \pt for any of the analyzed event samples.
In the case of the SD-enhanced event sample, the predictions of \PYTHIA{}8 \CUETM provide the best description of the data, while those of \EPOS\ \LHC disagree by up to $\approx$40\%.

For the distribution of the integrated leading charged particle \PT as a function of \ptmin, the predictions are normalized to the data in the high-\ptmin region, since this region is  better described by the models.
The \ptmin distribution for the SD-enhanced event sample is very different from the others, and the normalization  is performed at \mbox{$\ptmin=3.2\GeV$},  while for inelastic and NSD-enhanced samples it is performed at \mbox{$\ptmin = 9 \GeV$}.
The inelastic and NSD-enhanced event samples are best described by the predictions of \EPOS\ \LHC and \PYTHIA{}8 \CUETM{}, although the former overestimates particle production by about \mbox{$ 10\%$} at around 4--5\GeV, and the latter underestimates it by a similar amount at around \mbox{$\ptmin =1\GeV$}.
The predictions of \PYTHIA{}8 \MBR agree with the data  in the high-\pt region above {9}{\GeV}
but increasingly underestimate the data at lower \pt\ values,
 where discrepancies of up to about $20\%$ are observed.
The predictions of \PYTHIA{}8 \CUETM describe best the SD-enhanced data set, while those of \EPOS\ \LHC and \PYTHIA{}8 \MBR overestimate and underestimate the data by up to about $40\%$, respectively.
Comparing the shapes of the  $D(\ptmin)$ distributions for the inelastic (or NSD-enhanced) and
SD-enhanced samples, the transition between the regions dominated by particle production from MPIs
(and softer diffractive scatterings) and from single-hard parton scatterings seemingly occurs at
about $4 \GeV$ and $2 \GeV$, respectively, as indicated by the (fast) change of slope in the spectra
around these \ptmin values.

\section{Summary}
\label{sec:sum}
Charged particle distributions measured with the CMS detector in  minimum bias  proton-proton collisions at a center-of-mass energy of $\sqrt{s}=13\TeV$ have been presented.
Charged particles are selected with transverse momenta satisfying  \mbox{$\pt>0.5\GeV$} in the pseudorapidity range $\abs{\eta} < 2.4$.
The measured distributions, corrected for detector effects, are presented for three different event samples selected according to the maximum particle energy in the range $3 < \abs{\eta} < 5$.
The event samples correspond to an inelastic sample, a sample dominated by nonsingle diffractive dissociation events (NSD-enhanced sample), and an event sample enriched by single diffractive dissociation events (SD-enhanced sample).
 \begin{tolerant}{800}
In general, the  event generators \EPOS\ \LHC{}, \PYTHIA{}8 \CUETM, and \PYTHIA{}8 \MBR
describe the measurements reasonably well.
However, differences are observed in the \pse distributions for the SD-enhanced event sample in the region where the diffractive dissociative system is observed.
In the distributions integrated over the \pt  of the leading particle above a given threshold,
$D(\ptmin)$, deviations of up to 40\% are observed in the small \pt region.
The change from a relatively flat to a falling $D(\ptmin)$ distribution occurs at different \ptmin values for the diffractive-enhanced  event samples ($\ptmin \approx2$\GeV) and the inelastic and NSD-enhanced sample  ($\ptmin  \approx4$\GeV).
\end{tolerant}
The level of agreement between these new measurements at  $\sqrt{s}=13\TeV$  and the event generators predictions is comparable to that observed for previous measurements at lower energies.
The measurements described here provide new insights into
low momentum-exchange parton scatterings
that dominate inelastic (including diffractive)  \Pp\Pp\ interactions.
The rich variety of distributions presented for different event samples, especially those enhanced in diffractive processes, provide new information to understand the transition from  perturbative to  nonperturbative regions in particle production in high-energy  \Pp\Pp\   collisions and help constrain model parameters in modern hadronic event generators  used in collider and cosmic-ray physics.

\begin{acknowledgments}
We congratulate our colleagues in the CERN accelerator departments for the excellent performance of the LHC and thank the technical and administrative staffs at CERN and at other CMS institutes for their contributions to the success of the CMS effort. In addition, we gratefully acknowledge the computing centers and personnel of the Worldwide LHC Computing Grid for delivering so effectively the computing infrastructure essential to our analyses. Finally, we acknowledge the enduring support for the construction and operation of the LHC and the CMS detector provided by the following funding agencies: BMWFW and FWF (Austria); FNRS and FWO (Belgium); CNPq, CAPES, FAPERJ, and FAPESP (Brazil); MES (Bulgaria); CERN; CAS, MoST, and NSFC (China); COLCIENCIAS (Colombia); MSES and CSF (Croatia); RPF (Cyprus); SENESCYT (Ecuador); MoER, ERC IUT, and ERDF (Estonia); Academy of Finland, MEC, and HIP (Finland); CEA and CNRS/IN2P3 (France); BMBF, DFG, and HGF (Germany); GSRT (Greece); OTKA and NIH (Hungary); DAE and DST (India); IPM (Iran); SFI (Ireland); INFN (Italy); MSIP and NRF (Republic of Korea); LAS (Lithuania); MOE and UM (Malaysia); BUAP, CINVESTAV, CONACYT, LNS, SEP, and UASLP-FAI (Mexico); MBIE (New Zealand); PAEC (Pakistan); MSHE and NSC (Poland); FCT (Portugal); JINR (Dubna); MON, RosAtom, RAS, RFBR and RAEP (Russia); MESTD (Serbia); SEIDI, CPAN, PCTI and FEDER (Spain); Swiss Funding Agencies (Switzerland); MST (Taipei); ThEPCenter, IPST, STAR, and NSTDA (Thailand); TUBITAK and TAEK (Turkey); NASU and SFFR (Ukraine); STFC (United Kingdom); DOE and NSF (USA).

\hyphenation{Rachada-pisek} Individuals have received support from the Marie-Curie program and the European Research Council and Horizon 2020 Grant, contract No. 675440 (European Union); the Leventis Foundation; the A. P. Sloan Foundation; the Alexander von Humboldt Foundation; the Belgian Federal Science Policy Office; the Fonds pour la Formation \`a la Recherche dans l'Industrie et dans l'Agriculture (FRIA-Belgium); the Agentschap voor Innovatie door Wetenschap en Technologie (IWT-Belgium); the F.R.S.-FNRS and FWO (Belgium) under the "Excellence of Science - EOS" - be.h project n. 30820817; the Ministry of Education, Youth and Sports (MEYS) of the Czech Republic; the Council of Science and Industrial Research, India; the HOMING PLUS program of the Foundation for Polish Science, cofinanced from European Union, Regional Development Fund, the Mobility Plus program of the Ministry of Science and Higher Education, the National Science Center (Poland), contracts Harmonia 2014/14/M/ST2/00428, Opus 2014/13/B/ST2/02543, 2014/15/B/ST2/03998, and 2015/19/B/ST2/02861, Sonata-bis 2012/07/E/ST2/01406; the National Priorities Research Program by Qatar National Research Fund; the Programa Severo Ochoa del Principado de Asturias; the Thalis and Aristeia programmes cofinanced by EU-ESF and the Greek NSRF; the Rachadapisek Sompot Fund for Postdoctoral Fellowship, Chulalongkorn University and the Chulalongkorn Academic into Its 2nd Century Project Advancement Project (Thailand); the Welch Foundation, contract C-1845; and the Weston Havens Foundation (USA).
\end{acknowledgments}

\newpage
\bibliography{auto_generated}

\cleardoublepage \appendix\section{The CMS Collaboration \label{app:collab}}\begin{sloppypar}\hyphenpenalty=5000\widowpenalty=500\clubpenalty=5000\vskip\cmsinstskip
\textbf{Yerevan Physics Institute, Yerevan, Armenia}\\*[0pt]
A.M.~Sirunyan, A.~Tumasyan
\vskip\cmsinstskip
\textbf{Institut f\"{u}r Hochenergiephysik, Wien, Austria}\\*[0pt]
W.~Adam, F.~Ambrogi, E.~Asilar, T.~Bergauer, J.~Brandstetter, E.~Brondolin, M.~Dragicevic, J.~Er\"{o}, A.~Escalante~Del~Valle, M.~Flechl, R.~Fr\"{u}hwirth\cmsAuthorMark{1}, V.M.~Ghete, J.~Hrubec, M.~Jeitler\cmsAuthorMark{1}, N.~Krammer, I.~Kr\"{a}tschmer, D.~Liko, T.~Madlener, I.~Mikulec, N.~Rad, H.~Rohringer, J.~Schieck\cmsAuthorMark{1}, R.~Sch\"{o}fbeck, M.~Spanring, D.~Spitzbart, A.~Taurok, W.~Waltenberger, J.~Wittmann, C.-E.~Wulz\cmsAuthorMark{1}, M.~Zarucki
% ...
\vskip\cmsinstskip
% ...
1:  Also at Vienna University of Technology, Vienna, Austria\\
% ...
\end{sloppypar}
\end{document}